\documentclass[showpacs,aps,prb,twocolumn,superscriptaddress,10pt]{revtex4-2}

\usepackage[english]{babel}
\usepackage[utf8]{inputenc}
\usepackage[T1]{fontenc} 
\usepackage[normalem]{ulem}
\usepackage{amsmath,amssymb,amsfonts,physics,bm,graphicx,xcolor,mathtools,dsfont}
\usepackage{mathptmx}
\usepackage[colorlinks=true,linkcolor={blue!65!red},citecolor={blue!65!red},urlcolor={blue!60!black}]{hyperref}
\usepackage[nameinlink]{cleveref}
\usepackage{mathrsfs}
\usepackage{xcolor}
\usepackage{orcidlink}
\usepackage{xspace}
\usepackage{upgreek}

\newcommand{\mLLG}{os-LLG\xspace}
\newcommand{\alphaG}{\alpha_\mathrm{G}\xspace}
\newcommand{\alp}{\mathcal{A}\xspace}
\newcommand{\alpham}{\alpha_{\mathrm{m}}\xspace}
\newcommand{\betam}{\beta_{\mathrm{m}}\xspace}

\hypersetup{
    breaklinks=true,
    colorlinks=true,
    urlcolor=blue,
    linkcolor=blue,
    citecolor=blue}
    
\let\oldbibitem\bibitem 
\renewcommand{\bibitem}{
    \renewcommand{\doi}[1]{\texttt{\href{https://doi.org/##1}{doi:##1}}} 
    \let\bibitem\oldbibitem 
    \oldbibitem 
}

\begin{document}

\title{Atomistic spin dynamics with quantum colored noise}

\author{Fried-Conrad Weber\orcidlink{0009-0002-5135-0249}}
\email{fried-conrad.weber@uni-potsdam.de}
\affiliation{Helmholtz-Zentrum Berlin für Materialien und Energie GmbH, Wilhelm-Conrad-Röntgen Campus,
BESSY II, 12489 Berlin, Germany}
\affiliation{University of Potsdam, Institute of Physics and Astronomy, Karl--Liebknecht--Str. 24--25, 14476 Potsdam, Germany}
\author{Felix Hartmann\orcidlink{0009-0006-6048-175X}}
\affiliation{University of Potsdam, Institute of Physics and Astronomy, Karl--Liebknecht--Str. 24--25, 14476 Potsdam, Germany}
\author{Matias Bargheer\orcidlink{0000-0002-0952-6602}}
\affiliation{Helmholtz-Zentrum Berlin für Materialien und Energie GmbH, Wilhelm-Conrad-Röntgen Campus,
BESSY II, 12489 Berlin, Germany}
\affiliation{University of Potsdam, Institute of Physics and Astronomy, Karl--Liebknecht--Str. 24--25, 14476 Potsdam, Germany}
\author{Janet Anders\orcidlink{0000-0002-9791-0363}}
\affiliation{University of Potsdam, Institute of Physics and Astronomy, Karl--Liebknecht--Str. 24--25, 14476 Potsdam, Germany}
\affiliation{Department of Physics and Astronomy, University of Exeter, Stocker Road, Exeter EX4 4QL, UK}
\author{Richard F L Evans\orcidlink{0000-0002-2378-8203}} 
\email{richard.evans@york.ac.uk} 
\affiliation{School of Physics, Engineering and Technology,
University of York, York YO10 5DD, UK}

\begin{abstract}
The accurate prediction of temperature-dependent magnetization dynamics is a fundamental challenge in computational magnetism. While Atomistic Spin Dynamics (ASD) simulations have emerged as a powerful tool for studying magnetic phenomena, their classical nature leads to significant deviations from experimental observations, particularly at low temperatures. Here we present a comprehensive implementation of quantum-corrected ASD into the \textsc{vampire} software package, based on the open-system Landau-Lifshitz-Gilbert equation with a quantum thermostat. Our implementation incorporates memory effects along with colored noise derived from quantum-mechanical considerations that improve the description of the equilibrium magnetization. We demonstrate excellent quantitative agreement with experimental magnetization curves for nickel and gadolinium across the full temperature range. 
Our results establish that incorporating quantum environmental effects and colored noise substantially enhances the predictive capabilities of ASD simulations, providing a robust framework for modeling temperature-dependent magnetic phenomena in localized moment magnetic systems.
\end{abstract}

\maketitle


\section{Introduction}
Atomistic spin models are foundational tools for investigating Heisenberg-like ferromagnets, offering detailed insights into magnetic properties arising from localized moments in lattice structures. These models, which describe magnetism through classical localized magnetic moments on a lattice coupled by exchange interactions, have successfully captured the properties of diverse magnetic systems—from binary alloys like FePt~\cite{Mryasov_2005} and GdFe~\cite{Radu_2011} to more complex materials such as Fe$_3$O$_4$~\cite{Nedelkoski_2017}, IrMn~\cite{JenkinsPRM2021}, FeRh~\cite{Barker_2015}, YIG~\cite{Barker_2016}, and Nd$_2$Fe$_{14}$B~\cite{Toga_2016, Miyashita_2018}. Atomistic spin dynamics (ASD) simulations have proven invaluable for understanding ultrafast magnetization dynamics, enabling detailed studies of phenomena such as thermal fluctuation effects on Gilbert damping~\cite{Sampan-a-paiPRA2019}, sub-picosecond laser-induced demagnetization in Ni~\cite{Evans_2015}, and thermally driven switching in GdFe~\cite{Radu_2011}. Recent advancements have expanded the capabilities of ASD models through developments like Spin-Lattice Dynamics (SLD), which incorporates lattice vibrations~\cite{Strungaru_2021, Ma_2008, Tranchida_2018, Muller_2019, Hellsvik_2019, Lange_2023}; Longitudinal-Spin Fluctuations (LSF), capturing Stoner-like effects in transition metals~\cite{Ma_2012,EllisPRB2019}; and high and multi-temperature frameworks describing energy flow between electronic, magnetic, and phononic subsystems following ultrafast laser excitation~\cite{Zahn_2021, Pankratova_2022}. These innovations have broadened the scope of ASD simulations to address increasingly complex magnetic phenomena, with LSF playing a particularly significant role for modeling itinerant magnetic materials~\cite{Czirjak_2023}.

However, ASD simulations face fundamental limitations due to their classical treatment of magnetic moments via the Landau-Lifshitz-Gilbert (LLG) equation with Gaussian white noise. This classical approach inadequately captures the temperature-dependent magnetization, especially near absolute zero, where quantum effects become significant. Recent advances have addressed these limitations through complementary approaches. Barker and Bauer~\cite{Barker_2019} implemented a semi-quantum thermostat that introduces correlated noise obeying quantum statistics without zero-point fluctuations, enabling accurate modeling of low-temperature magnetization in complex materials. Bergqvist and Bergman~\cite{BergqvistPRM2018} developed a method incorporating the magnon density of states to apply quantum (Bose-Einstein) statistics in atomistic simulations, significantly improving the prediction of thermodynamic properties. More recently, Berritta \textit{et al}~\cite{Berritta_2024} provided an analytical framework for temperature rescaling between classical and quantum regimes. Nevertheless, a fully microscopic model capable of accurately reproducing the temperature dependence of magnetization across all regimes has remained elusive.

In this paper, we present an implementation of atomistic spin dynamics using an open-system Landau-Lifshitz-Gilbert (\mLLG) equation derived from a microscopic open quantum system approach with colored noise~\cite{Anders_2022}. To validate the method, we compare the paramagnetic case (exchange constant $J_{ij} = 0$) with results from \textsc{SpiDy}~\cite{Scali_2024}, a software package designed for solving the \mLLG~equation with colored noise for single and small clusters of spins. The integration of the \mLLG~equation into \textsc{vampire}, a C\texttt{++}-based software package for ASD simulations~\cite{Evans_2014} enables us to study the temperature-dependent magnetization of elementary room temperature magnetic materials, including Ni, Fe, Co, and Gd. Our findings demonstrate that atomistic \mLLG~simulations with a colored semi-quantum (which we will call ``quantum no-zero'') thermostat significantly improve the accuracy of temperature-dependent magnetization predictions for Ni and Gd. The model's Heisenberg-like nature limits its applicability for materials such as Co and Fe. This paper aims to underscore the potential of \mLLG~simulations in advancing the study of temperature-dependent magnetic phenomena, while also highlighting their current limitations and areas for future development.
\begin{figure*}
    \centering
    \includegraphics[width=\textwidth]{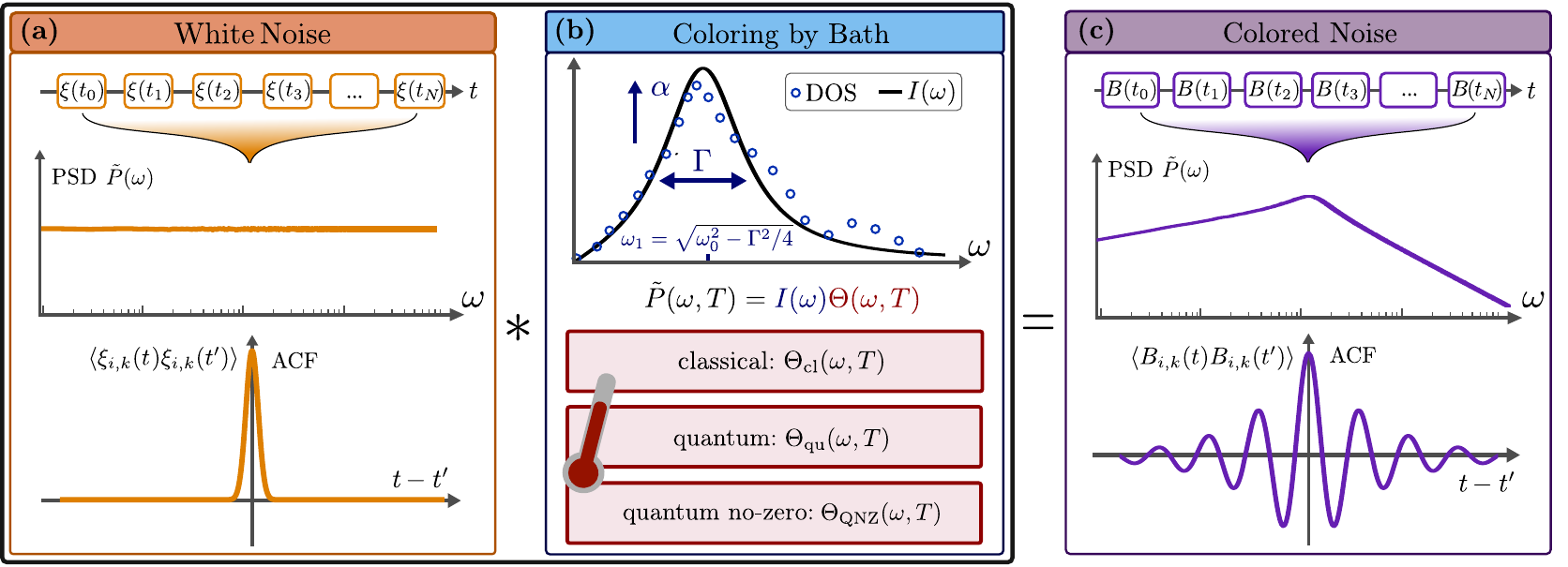}
    \caption{\textbf{Colored noise generation:} In this Figure we sketch schematically how the colored noise is generated (compare with Sec.~\ref{sec:numimplementation}).
    (a) White noise is generated by drawing $N$ random numbers $\xi$ from a Gaussian probability distribution $p(\xi)$. 
    The white noise is characterized by a flat power spectral density (PSD) and a $\delta$-shaped autocorrelation function (for better visibility approximated by a slim Gaussian distribution).  
    (b) To color the noise the power spectrum needs to be fixed: Firstly, the magnetic environment is approximated by a Lorentzian spectral density $I^{\mathrm{Lor}}(\omega)$ (black solid curve, middle panel). The parameters $(\alp,\Gamma,\omega_0)$ are obtained by fitting $I^{\mathrm{Lor}}(\omega)$ to the (phonon) density of states DOS (blue points). 
    Secondly, one of three thermostats $\Theta$ (classical, quantum, quantum no-zero) must be selected.
    (c) Via the convolution in Eq.~\eqref{eq:noise_convolution} the noise becomes colored, as the frequencies are weighted and the Lorentzian introduces a natural high-frequency cut-off.
    The colored noise $B_{\mathrm{th},i}^k(t)$ has a shaped powers spectrum~\eqref{eq:PSD_definition} and a finite autocorrelation function. 
    }
    \label{fig:num_implementation} 
\end{figure*}
\section{Theory}
\label{sec:theory}

Unlike the phenomenological Landau-Lifshitz-Gilbert equation widely used in atomistic spin dynamics simulations, the open-system Landau-Lifshitz-Gilbert (\mLLG) equation emerges naturally from an open quantum system approach~\cite{Anders_2022}. This fundamental derivation provides a more rigorous theoretical foundation for describing the dynamics of classical spin vectors interacting with a thermal bosonic bath. 
Historically the temperature dependence of magnetization is described by Bloch's Law $m(T) \sim m_0 T^{3/2}$, and attributed to the presence of quantized spin waves known as magnons~\cite{Bloch1930}. As quantum quasiparticles with finite energy and spin-1 (due to the flip of a single spin from $+^1/_2 \hbar \rightarrow -^1/_2\hbar$), these are assumed to be bosons and thus follow Bose-Einstein statistics. In textbooks and other works~\cite{Sun2025,Chen2023,Hogg2024} this is extended to assume that the thermal bath itself is also bosonic, but a consideration of the origin of spin fluctuations suggests an important contributions from the electron system, especially for ultrafast dynamics~\cite{Born2021}. This suggests that for metallic magnets the relevant heat bath is likely fermionic which may have important effects at short timescales, and require modifications to the derivation of the quantum thermostat~\cite{Anders_2022}. In this work we primarily consider the quasi-equilibrium regime where our magnetic system is coupled to a bosonic heat bath reflecting lattice fluctuations. 

The derived \mLLG~equation for the $i$'th spin $\mathbf{S}$ of Ref.~\cite{Anders_2022} is
\begin{equation}
    \dot{\mathbf{S}}_i(t) = \gamma\,\mathbf{S}_i(t)\times\left[ \mathbf{B}_{\mathrm{eff},\,i} + \mathbf{B}_{\mathrm{th},\,i}(t) + \int_{t_0}^{t} \mathrm{d}t' K(t-t')\mathbf{S}_i(t') \right].
    \label{eq:mLLG}
\end{equation}
The first contribution to Eq.~\eqref{eq:mLLG} is a time-independent effective field $\mathbf{B}_{\mathrm{eff},\,i}$ that combines an applied field $\mathbf{B}_{\mathrm{app}}$, an anisotropy field $\mathbf{B}_{\mathrm{ani}}$, and an exchange field $\mathbf{B}_{\mathrm{exc}}$, i.e. $\mathbf{B}_{\mathrm{eff},\,i}= \mathbf{B}_{\mathrm{app}} + \mathbf{B}_{\mathrm{ani}} + \mathbf{B}_{\mathrm{exc},\,i}$.
The second term in Eq.~\eqref{eq:mLLG} refers to a time-dependent stochastic thermal field $\mathbf{B}_{\mathrm{th},i}(t)$, which arises from coupling each spin to a bosonic bath (e.g. phonons)~\cite{Anders_2022,Scali_2024}. The third and final contribution to Eq.~\eqref{eq:mLLG} is a memory kernel term $K(t-t')$ that introduces non-Markovian dynamics by making the spin evolution dependent on its history. It contains the damping of the system and takes inertial corrections into account.
The noise term and the memory kernel are related self-consistently by the fluctuation-dissipation theorem~\cite{Anders_2022}.

In general, the time correlations in the thermal noise are determined by setting the coupling between the spin system and the bath via a power spectrum 
\begin{equation}
    \tilde{P}(\omega,T) = \frac{1}{2}\int_{-\infty}^{\infty}\mathrm{d}t'\,e^{i\omega t'}\langle B^{k}_{\mathrm{th},i}(t)B^{l}_{\mathrm{th},i}(t') \rangle_\beta,
    \label{eq:PSD_definition}
\end{equation}
where $\langle\bullet\rangle_\beta$ indicates averaging with respect to the thermal distribution. 
In the following our power spectrum is of Lorentzian shape, i.e. $\tilde{P}^{\mathrm{Lor}}(\omega,T) = I^{\mathrm{Lor}}(\omega)\Theta(\omega,T)$, with
\begin{align}
    {I}^\mathrm{Lor}(\omega) =  \frac{\alp\Gamma\hbar\omega}{(\omega_0^2-\omega^2)^2 + \omega^2\Gamma^2},
    \label{eq:powerspectrum}
\end{align}
where $\omega_0$ represents the center frequency, $\Gamma$ the width, and $\alp$ the amplitude (coupling strength) of the Lorentzian spectral density $I^{\mathrm{Lor}}(\omega)$. Here the Lorentzian form introduces a characteristic resonance frequency of the bath $\omega_0$ that couples to the spin system, but with a defined width $\Gamma$ arising from phonon dispersion. The coupling strength $\alp$ defines the strength of coupling between the bath and the spin system and is directly related to the Gilbert damping parameter $\alphaG$ and the spin-orbit coupling.
Choosing a Lorentzian spectral density introduces a spectral character to the effective power spectrum, that promotes resonant thermal excitations at $\omega = \omega_0$, while depopulating the available excitations at higher frequencies $\omega > \omega_0$ (for a detailed discussion see App.~\ref{app:ohmic_vs_Lorentzian_PSD}).

The temperature dependence of the power spectrum $\tilde{P}^{\mathrm{Lor}}(\omega,T)$ is introduced by one of the following three distinct thermostats $\Theta(\omega,T)$: 
\begin{align}
    \Theta_{\mathrm{cl}}(\omega,T) &= \frac{2 k_{\mathrm{B}}T}{\hbar\omega}, \label{eq:classical} \\
    \Theta_{\mathrm{qu}}(\omega, T) &= \coth\left(\frac{\hbar\omega}{2k_{\mathrm{B}}T}\right), \label{eq:quantum} \\
    \Theta_{\mathrm{qnz}}(\omega, T) &= \coth\left(\frac{\hbar\omega}{2k_{\mathrm{B}}T}\right)-1. \label{eq:barker}
\end{align}
Eq.~\eqref{eq:classical} represents the classical thermostat~\cite{Evans_2014}, Eq.~\eqref{eq:quantum} the quantum thermostat~\cite{Anders_2022}, and Eq.~\eqref{eq:barker} the quantum no-zero (QNZ) thermostat~\cite{Barker_2019}. The quantum thermostat includes zero-point fluctuations at $T = 0$~K, while both the classical and QNZ thermostats ensure vanishing fluctuations at absolute zero. In the high-temperature limit $\hbar \omega \gg 2k_\text{B} T$ the quantum noise converges to the classical behavior.

While in this paper we solely consider a Lorentzian PSD for the noise (where $I(\omega)$ takes the form of Eq.~\eqref{eq:powerspectrum}), it is also possible to consider the same equation of motion (Eq.~\eqref{eq:mLLG}) and temperature dependencies (Eqs.~\eqref{eq:classical}-\eqref{eq:barker}) with Ohmic noise (so called due to the white noise nature of thermal noise in an ideal resistor). There the spectral density of the bath is linear in frequency, i.e. $I^{\mathrm{Ohm}}(\omega) \propto \alphaG\omega$. In this case, Eq.~\eqref{eq:mLLG} reduces to the standard LLG equation~\cite{Anders_2022} and for a classical thermostat (Eq.~\eqref{eq:classical}) the thermal fluctuations lose their time-correlated nature. 
Our approach allows for a number of different \textit{flavors} of thermal bath within our open-system framework, summarized in Tab.~\ref{tab:taxonomy}. Standard classical spin dynamics follows a classical thermostat with Ohmic (white) noise. The additional flexibility of this frameworks allows independent consideration of the effects of time correlation and quantum statistics, with and without zero-point fluctuations.

\begin{table}[!tb]
\centering 
\caption{\textbf{Taxonomy of colored thermostats}, including the Ohmic and Lorentzian bath spectral densities $I(\omega)$, each with classical, quantum, and quantum no-zero thermostats $\Theta(\omega,T)$.}
\begin{ruledtabular}
\begin{tabular}{l c c c} 
PSD $\tilde{P}(\omega,T)$ & Acronym & $I(\omega)$ & $\Theta(\omega,T)$ \\
\hline 
classical noise & CO & Ohmic &  $T/\omega$\\
quantum Ohmic & QO & Ohmic & $\coth(\omega/T)$ \\
quantum no-zero Ohmic & QNZO & Ohmic & $\coth(\omega/T)-1$\\
classical Lorentzian & CL & Lorentzian & $T/\omega$\\
quantum Lorentzian   & QL & Lorentzian & $\coth(\omega/T)$ \\
quantum no-zero Lorentzian & QNZL & Lorentzian & $\coth(\omega/T)-1$ \\
\end{tabular}
\end{ruledtabular}

\label{tab:taxonomy}
\end{table}

The Lorentzian-shaped spectrum $I^{\mathrm{Lor}}(\omega)$ in Eq.~\eqref{eq:powerspectrum} is used to approximate the physical bath, which is characterized by the phonon density of states~\cite{Nemati_2022}.
The arising noise field $\mathbf{B}_{\mathrm{th},\,i}$ is characterized by $\langle \mathbf{B}_{\mathrm{th},\,i}(t)\rangle = 0$ and its temporal autocorrelation is given by $\langle B_{\mathrm{th},i}^{k}(t) B_{\mathrm{th},j}^{l}(t') \rangle \propto \delta_{ij}\delta_{kl}\chi(t-t',T)$, where $\chi(\cdot)$ depends on the chosen parameters of Eq.~\eqref{eq:powerspectrum}. The indices $k$ and $l$ denote the different spatial direction of the noise field vector $\mathbf{B}_{\mathrm{th},\,i}$, i.e. $x$, $y$, and $z$. In general, $\chi(t-t',T)$ is not delta-correlated, leading to colored noise (compare with Fig.~\ref{fig:num_implementation}).

To fix the three Lorentzian parameters $\omega_0$, $\Gamma$, and $\alp$, we obtain $\omega_0$ and $\Gamma$ by fitting the Lorentzian spectral density $I^{\mathrm{Lor}}(\omega)$ to the phonon density of states (DOS)~\cite{Nemati_2022}, representing the long-time fluctuations of the heat bath. In this approximation, we assume that the electronic system is always in its ground state and thermal fluctuations arise due to fluctuations of the lattice. The last free parameter $\alp$ is fixed by comparing it to the (Gilbert) damping parameter of the material, via $\alphaG = \alp\Gamma/\omega_0^4$~\cite{Anders_2022}. More details of the fitting are given in Appendix~\ref{app:fitDOS}.

In summary, the two advantages of the open-system LLG equation~\eqref{eq:mLLG}, compared to the standard LLG equation, are that it contains colored quantum thermostats and introduces non-Markovian effects in a thermodynamically consistent way.
The Lorentzian power spectrum introduces a natural cut-off frequency at finite frequency, so that the characteristic system frequencies (given here by the phonons) are weighted and the noise is colored. 
Generally, colored noise with different thermostats and memory effects are considered to play a crucial role at short times on the microscale~\cite{Pershin_2011,Atxitia_2009}, but here we also consider the effects on the equilibrium magnetization.

\section{Numerical Implementation}
\label{sec:numimplementation}
To translate the theoretical advantages of a quantum thermostat with colored noise described by Eq.~\ref{eq:mLLG} into practical ASD simulations we follow the numerical methods outlined in Refs.~\cite{Anders_2022,Scali_2024} and implement these into the \textsc{vampire} software package~\cite{Evans_2014}. Below we outline our approach for generating colored noise and implementing non-Markovian spin dynamics into ASD as schematically illustrated in Fig.~\ref{fig:num_implementation}.

Before the noise is colored, white noise (with uncorrelated random fluctuations) is generated by drawing $N$ random numbers $\xi^k_i$ per spatial dimension $k$ from a Gaussian probability distribution $p(\xi^k_i)$, with mean $\mu = 0$ and variance $\sigma = 1/\sqrt{\mathrm{d} t}$, where $\mathrm{d} t$ is the integration step-size (see Fig.~\ref{fig:num_implementation}(a)). 
The white noise time-signal $\xi^k_i(t)$ of spatial dimension $k$ is Fourier transformed into frequency-space, such that $\tilde{\xi}^k_i(\omega)$ is obtained. The Fourier transformed noise is then scaled by multiplying by the square root of the Lorentzian power spectrum $\tilde{P}^{\mathrm{Lor}}(\omega,T)$ (Eq.~\eqref{eq:powerspectrum}) including the appropriate thermostat, see Tab.~\ref{tab:taxonomy} to color the noise, shown schematically in Fig.~\ref{fig:num_implementation}(b). Finally the colored noise is inverse Fourier transformed returning the noise to the time domain, expressed mathematically as the integral over frequency
\begin{equation}
    B_{\mathrm{th},\,i}^k(t) = \int_{-\infty}^{\infty} \frac{\mathrm{d}\omega}{2\pi} e^{-\text{i}\omega t} \sqrt{\tilde{P}^{\mathrm{Lor}}(\omega, T)} \tilde{\xi}^k_i(\omega).
    \label{eq:noise_convolution}
\end{equation}
The resulting noise is colored and possesses a finite autocorrelation in time that is characteristic of the chosen Lorentzian parameters, thermostat and temperature. 
Numerically we utilize the FFTW library to compute the Fourier transforms~\cite{Frigo2005} efficiently in $N \log N$ time, where $N$ is the number of elements. 

The above procedure requires generating the whole noise for the simulation at the start, with three components of noise for each atom and one value for each timestep. For a small sized atomistic system of 10,000 atoms, 1 ns simulation time and $\mathrm{d}t_{\text{sim}} = 0.1$ fs timestep, this requires 2.4 TB of memory not including data for the Fourier transforms and so is not typically available in today's desktop computers. We therefore introduce a number of optimizations to improve the memory efficiency of the implementation.

Firstly, the Fourier transformation does not need to be evaluated with the small time steps required for simulating the magnetization dynamics. This is due to the fact that the colored noise varies on timescales of 1-100 fs. Thus the noise can be generated on a coarse time grid with a timestep $\mathrm{d}t_{\text{coarse}} = D \cdot \mathrm{d}t_{\text{sim}}$, where $D$ is a decimation factor determined by the characteristic frequencies of the noise. We select $D$ such that $\mathrm{d}t_{\text{coarse}}$ satisfies the Nyquist criterion for the highest significant frequency component in the PSD, typically $D \approx 10-100$ for our Lorentzian parameters. The noise values at intermediate simulation timesteps are then obtained through linear interpolation. To preserve the correct statistical properties, we scale the initial white noise by $\sqrt{\mathrm{d}t_{\text{sim}}/\mathrm{d}t_{\text{coarse}}}$, which ensures that the standard deviation of the interpolated noise and the autocorrelation function matches that of noise generated directly at the fine timestep. This approach reduces both memory requirements and computational costs by a factor of $D$, making it feasible to generate colored noise for large-scale simulations while maintaining the essential statistical properties that influence the system dynamics.

Secondly, the number of noise sources does not strictly have to be the same as the number of atoms due to the fact each atom has three independent components and the fact that the noise is rapidly varying in time. We can therefore set a much smaller number of noise components, typically 20\% of the number of atoms, and randomly assign these to each atom and component $k$ with a random three-dimensional rotation to avoid spatial correlations. This further reduces the memory requirements by a factor 5. The combination of these optimizations makes larger-scale simulations feasible. We also parallelize our implementation alongside the \textsc{Vampire} parallelization using the message passing interface (MPI) with shared memory buffers to efficiently generate the noise in a multiprocessor system, leading to near linear speedup of the execution time with the number of processors.

So far we have considered the generation of the thermal noise with the desired PSD and thermodynamics, but we have not considered exactly how to solve Eq.~\ref{eq:mLLG} including the memory kernel $K(t-t')$. Explicitly this would require an integration over all time to include the memory effects and the associated dissipation so that the fluctuation-dissipation theorem is satisfied. 
Previously it has been shown in Ref.~\cite{Anders_2022}, that under the assumption of Lorentzian bath spectrum in~\eqref{eq:PSD_definition}, Eq.~\eqref{eq:mLLG} can be recast into a set of three coupled differential equations. Herefore, two auxiliary variables $\mathbf{V}$ and $\mathbf{W}$~\cite{Scali_2024,Anders_2022} are needed to capture the non-Markovian nature of the \mLLG~equation, given by:
\begin{align}
    \dot{\mathbf{S}}_i(t) &= \gamma\,\mathbf{S}_i(t)\times\left( \mathbf{B}_{\mathrm{eff},\,i} + \frac{1}{\sqrt{S_0}}\mathbf{B}_{\mathrm{th},\,i}(t) + \mathbf{V}_i(t)\right),
    \label{eq:generalizedLLG_numerI}\\
    \dot{\mathbf{V}}_i(t) &= \mathbf{W}_i(t),  \label{eq:generalizedLLG_numerII}\\
    \dot{\mathbf{W}}_i(t) &= \alp\mathbf{S}_i(t) - \omega_0^2 \mathbf{V}_i(t) - \Gamma \mathbf{W}_i(t),
    \label{eq:generalizedLLG_numerIII}
\end{align}
where $S_0$ represents the length of the classical spin vector.  Here the kernel $K(t-t')$ represents the dissipation of the energy to the bath and is constructed to obey the fluctuation-dissipation theorem~\cite{Callen1951} in the general case where the noise is colored and has a different thermostat~\cite{Anders_2022}. 
The exchange field $\mathbf{B}_{\mathrm{exc},\,i}$ follows the Vampire framework \cite{Evans_2014}. 
The thermal fields $\mathbf{B}_{\mathrm{th},\,i}$ are precalculated for each atom and each spatial direction numerically by using the above method. 
Finally, Eqs.~\eqref{eq:generalizedLLG_numerI}-\eqref{eq:generalizedLLG_numerIII} are solved using a fourth-order Runge-Kutta (RK4) integrator, which computes four intermediate slopes ($k_1$–$k_4$) to update the state:
\begin{align*}
    y_{+1} = y_n + \frac{\mathrm{d} t}{6} \left(k_1 + 2 k_2 + 2 k_3 + k_4 \right).
\end{align*}
Because the exchange field depends sensitively on the instantaneous spin configuration, it is updated and recalculated at each intermediate RK4 step based on the spins’ predicted positions. The precomputed thermal noise field is incorporated using the previously described time-interpolation with decimation: for the first RK4 slope ($k_1$) we use the noise at time $t_n$, for the intermediate slopes ($k_2$ and $k_3$) we use the average of the noise at $t_n$ and $t_{n+1}$, and for the final slope ($k_4$) we use the noise at time $t_{n+1}$. To validate our numerical implementation, we perform a consistency check in the paramagnetic regime $(J_{ij} = 0)$ with the single spin implementation~\textsc{SpiDy}~\cite{Scali_2024}. 
The results are detailed in App.~\ref{app:para_dynamics} and show excellent agreement between the two implementations.

\section{Numerical results}
\label{sec:numerical_results}
\begin{figure*}
    \centering
    \begin{centering}
    \includegraphics[width=\textwidth]{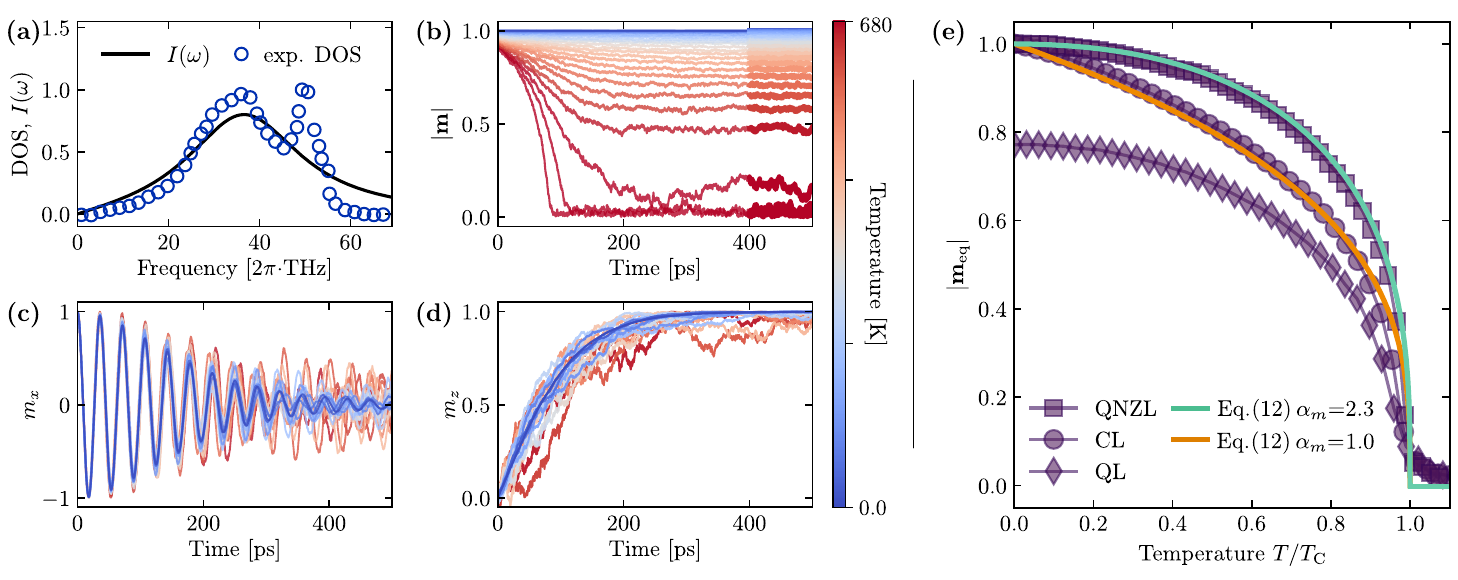}
    \caption{\textbf{Equilibrium magnetization of nickel using (quantum no-zero) colored noise (QNZL):
    } (a) Fit of the measured phonon DOS for nickel (open blue circles)~\cite{Kresch_2007} with the Lorentzian spectral density $I^{\mathrm{Lor}}(\omega)$ (black solid line) defining spin-bath coupling. Averaged dynamics of the (b) magnetization amplitude, (c) $m_x$ component, and (d) $m_z$ component, using QNZL noise for different temperatures (blue ($0$~K) to red ($680$~K) transition). 
    (e) Equilibrium magnetization for classical thermostat (circle), quantum thermostat (diamond), and quantum no-zero thermostat (squares). 
    While the \mLLG~simulation with the classical thermostat reproduces the classical ASD simulation results~\cite{Evans_2015} (i.e. Eq.~\eqref{eq:mT} with $\alpha_\mathrm{m} = 1$, orange solid line), the \mLLG~simulation with the QNZL noise matches equation ~\eqref{eq:mT} with $\alpha_\mathrm{m} = 2.32$ fitted for nickel (green solid line). 
    Simulations are for a $7\times7\times7$~nm block containing $45\,000$ spins.
    }
    \label{fig:equilibrium_noise}
    \end{centering}
\end{figure*}

To test our implementation in the multi-spin case we assume a simple spin Hamiltonian including nearest neighbor exchange and coupling to an external magnetic field of the form~\cite{Evans_2014}
\begin{equation}
    \mathscr{H} = -\sum_{i<j} J_{ij} \mathbf{S}_i\cdot \mathbf{S}_j - \mu_i \sum_i \mathbf{S}_i \cdot \mathbf{B}_\mathrm{app},
\end{equation}
where $J_{ij}$ is the exchange interaction between nearest neighbors, $\mathbf{S}_{i,j}$ represent local and neighboring atomic spin moment directions respectively, $\mu_i$ is the local spin moment and $\mathbf{B}_\mathrm{app}$ is the applied field vector. The system of coupled spins described by the Hamiltonian is then time-integrated using the \mLLG at constant temperature to achieve a consistent thermodynamic state. We first consider a simulation of Ni, with a low magnetic moment of $\mu_i = 0.61 \mu_\mathrm{B}$ and an intermediate case in terms of the strength of the quantum corrections. As described above in Section~\ref{sec:numimplementation} we characterize the bath by a Lorentzian spectral density $I^{\mathrm{Lor}}(\omega)$. The width $\Gamma = 28.9$~$\mathrm{rad}\cdot\mathrm{THz}$ and characteristic frequency $\omega_0 = 40.5$ $\mathrm{rad}\cdot\mathrm{THz}$ are obtained by fitting the phonon DOS of Ni, see Fig.~\ref{fig:equilibrium_noise}(a). 
Assuming that the Gilbert damping of Ni is $\alp = 4655~\mathrm{rad}^3\cdot\mathrm{THz}^3$, we obtain $\alphaG = 0.05$. We perform a simple computational experiment, initializing all spins parallel to the $x$-axis in a $B_z = 1.0$~T magnetic field and simulating the relaxation dynamics of the overall magnetization at different temperatures. Here we use a small system size of (7 nm)$^3$ with a face-centered cubic lattice containing around $N = 45\,000$ individual atomic spins. We use an exchange constant of $J_{ij} = 9.35 \times 10^{-22}$ J/link within the nearest neighbor approximation. All parameters for the simulations are summarized in Tab.~\ref{tab:parameters} and the dependency of the thermostat and the exchange energy on the Curie temperature are detailed in App.~\ref{app:Curie_temperature}.

Previous works have identified a clear difference comparing classical and quantum thermostats in magnetic systems in the temperature dependence of the magnetic order parameter $m(T)$ given by $m(T) = |\sum_i \mu_i\mathbf{S}_i|/\sum_i \mu_i$ where $\mu_i$ is the moment length. At $T = 0$ without zero-point motion all spins are aligned and give a fully saturated state. The shape of the $m(T)$ curve is characteristic of the thermodynamic fluctuations of the spins with increasing temperature, expressed by an interpolation of Bloch (low-temperature) and Curie (high-temperature) behavior and given by
\begin{equation}
    m(T) = \left[1-\left(\frac{T}{T_\mathrm{C}}\right)^{\alpham}\right]^{\betam},
    \label{eq:mT}
\end{equation}
where $\betam\simeq 0.34$ is the usual magnetization critical exponent describing the shape of the curve near the Curie temperature, and $\alpham$ is an exponent describing the shape of the curve at low temperature. This equation is a simplified form compared to that proposed by Kuz'min~\cite{Kuz_2005} but clearly separates classical and quantum behavior, allowing for cases where the system essentially follows the classical M(T) curve~\cite{SatoPRB2018}. The parameter $\alpham$ gives an indication of the \textit{quantumness} of the system, with $\alpham = 1$ the classical limit, and $\alpham > 1$ indicating quantum corrections, which are weak for Gd, strong for Fe, and intermediate for Co and Ni~\cite{Evans_2015}. The open-system LLG described in this paper for multi-spin systems naturally introduces quantum corrections to the spin fluctuations and we therefore use the temperature-dependent magnetization as a benchmark against experimental data to determine the accuracy of the parameters in our simulations.

The obtained averaged dynamics  $|\mathbf{m}(t)|$,  $m_x(t)$, and $m_z(t)$ for the QNZL noise at different temperatures is shown in Fig.~\ref{fig:equilibrium_noise}(b), (c), and panel (d), respectively. The system shows both precessional (Fig.~\ref{fig:equilibrium_noise}(c,d)) and longitudinal (Fig.~\ref{fig:equilibrium_noise}(b)) relaxation leading to a reduced magnetization length after around $200$ ps, after which an average equilibrium value of the magnetization is calculated. We note that this deviation from experimental measurements of ultrafast demagnetization likely occurs because we only include a phononic bath in our model; an electronic bath with higher frequency components would accelerate the equilibration timescale.

\begin{table*}
    \centering
    \caption{\textbf{Table of simulation parameters for the ferromagnetic elements Fe, Co, Ni and Gd} \cite{Evans_2014}. The Lorentzian parameters are obtained by fitting the phonon density of states for Fe \cite{Alp_2002}, Co \cite{Lizarraga2017}, Ni \cite{Kresch_2007} and Gd \cite{Li_2021}. Simulations use the atomistic spin moment as the spin length and not multiplies of $\hbar/2$. The exchange parameters for QNZL noise which reproduce the correct $T_\text{C}$ vary from the exchange parameters for classical ASD \cite{Evans_2014}. For our simulations, we do not consider the structural phase transitions of Fe and Co.}
    \begin{tabular}{lccccc}
    \hline
    \hline
    \textbf{}                                & Fe                    & Co               & Ni                  & Gd                     & Unit                   \\ \hline
Crystal structure                        & bcc                   & hcp   & fcc                   & hcp                    &                        \\
Unit cell size a                         & 2.866                 & 2.507 & 3.524                 & 3.636                  & \r{A}\\
Atomic spin moment $S_0$      & 2.22                  & 1.72  & 0.606                 & 7.63                   & $\mu_\mathrm{B}$       \\
Lorentzian amplitude $\alp$                 & 7812        & 4583                 & 4655          & 246  & $\mathrm{rad}^3\cdot\mathrm{THz}^3$ \\
Lorentzian width $\Gamma$               & 31.5                 & 27.1                 & 28.9                  & 10.0      & $\mathrm{rad}\cdot\mathrm{THz}$\\
Lorentzian center $\omega_0$            & 47.1                   &  39.7                & 40.5   & 14.9      & $\mathrm{rad}\cdot\mathrm{THz}$\\
Exchange energy $J_{\mathrm{ij}}$ (QNZL)     & $2.8 \times 10^{-21}$ &  $2.6 \times 10^{-21}$      & $9.35 \times 10^{-22}$& $5.35 \times 10^{-22}$& J/link                 \\ 
Curie temperature $T_\mathrm{C}$         & 1043                  & 1388  & 631                   & 293                    & K                      \\ \hline\hline
    \end{tabular}
    
    \label{tab:parameters}
\end{table*}

Starting with the classical thermostat and Lorentzian spectral density (CL noise), we find an excellent agreement with the standard classical behavior described by Eq.~\eqref{eq:mT} for $\alpham = 1$~(compare orange line and purple circles in Fig.~\ref{fig:equilibrium_noise}), confirming that the equilibrium magnetization of the open-system Landau-Lifshitz-Gilbert equation successfully recovers the standard stochastic LLG with white noise in the classical limit.

Next, the \mLLG equation with QL noise (Tab.~\ref{tab:taxonomy}) is simulated.
This provides crucial insights into the low-temperature regime (see purple diamonds in Fig.~\ref{fig:equilibrium_noise}(e)). While it accurately captures the shape of the magnetization curve at low temperatures, a distinctive feature emerges at absolute zero. 
The presence of quantum zero-point fluctuations, $\Theta_{\mathrm{qu}}(\omega,T=0~\mathrm{K}) \neq 0$, creates a competition with the ferromagnetic exchange coupling, resulting in a magnetization that remains below unity even at $T = 0$ K. Its origin is the fact that the thermodynamic fluctuations do not tend to zero in the quantum case, but plateau to a low temperature value that is temperature \textit{independent} (a more detailed discussion of the zero-point fluctuations can be found in App. \ref{app:Quantum_noise_disc}).

We further consider a form of the noise with the same quantum thermostat but backing off the zero-point fluctuations (QNZL), which removes the residual noise as $T \rightarrow 0$ leading to a fully saturated state at $m(0) = 1$. 
This is similar to the form used by Barker and Bauer~\cite{Barker_2016} where they used a quantum thermostat with an Ohmic power spectral density (QNZO).
The QNZL thermostat successfully reproduces the temperature dependence of magnetization of Ni across the entire temperature range, capturing both quantum effects at low temperatures and classical behavior near and above the Curie temperature, see purple squares in Fig.~\ref{fig:equilibrium_noise}(e). 
Our simulation results agree nicely with Eq.~\eqref{eq:mT} fitted from the experimental data for Ni (green line in Fig.~\ref{fig:equilibrium_noise}(e)).

For all thermostats we observe a slightly elevated magnetization in the paramagnetic phase near $T_\mathrm{C}$, which we attribute to finite size effects in our simulations (see Fig.~\ref{fig:equilibrium_noise}(e) for $T > T_{\mathrm{C}}$). 
This elevation is particularly noticeable given our relatively small sample size ($7\times7\times7$\, nm), though it does not affect the overall validity of our results in the ferromagnetic phase.

\begin{figure*}[!tb]
    \begin{centering}
    \includegraphics[width=\textwidth]{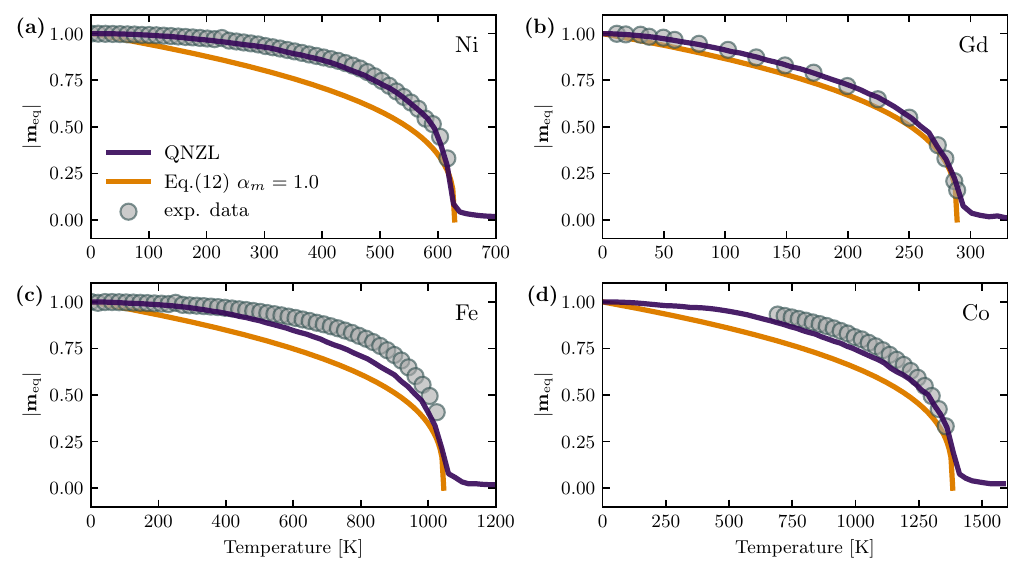}
    \caption{\textbf{Equilibrium magnetization for Ni, Gd, Fe, and Co:} Comparison of simulated equilibrium magnetization using the \mLLG~model with QNZL (purple solid line) and classical ASD simulations with classical white noise~\cite{Evans_2014} (orange solid line). The grey circles represent experimental data for (a) Nickel, (b) Gadolinium, (c) Iron~\cite{Crangle_1971}, (d) and Cobalt~\cite{Nigh_1963}. For Nickel and Gadolinium, the~\mLLG~model with QNZL noise shows excellent agreement with the measured temperature-dependent magnetization, while for Cobalt and Iron, there is a difference between the simulated and experimental results. Simulations performed for approximately 45\,000 spins with an effective field $B_z = 10$~mT. Further simulation parameters are detailed in Tab.~\ref{tab:parameters}. 
    }
    \label{fig:equilibrium}
    \end{centering}
\end{figure*}

Since the \mLLG~successfully reproduces the temperature-dependent equilibrium magnetization for Ni, we next consider comparative simulations for the elemental ferromagnets. In Fig.~\ref{fig:equilibrium}, we compare the equilibrium magnetization of the \mLLG~equation with QNZL noise (purple line) against classical ASD~\eqref{eq:mT} simulation results and experimental data (grey circles) for the four ferromagnetic materials: Nickel (a), Cobalt (b), Iron (c), and Gadolinium (d). Parameters for the different materials are given in Tab.~\ref{tab:parameters}.
Our simulations nicely recover the low temperature magnetization of all four materials, with excellent agreement for Nickel and Gadolinium over the full temperature regime, showing a strong improvement compared to the usual classical white noise ASD simulations. 

One important aspect of QNZ thermostats is that the shape (curvature) of the equilibrium magnetization curves is dependent on the length of the magnetic moment $S_0 := \mu_i$. This is different in standard ASD simulations with classical white noise (CO), where the shape of the curve is \textit{independent} of the magnetic moment~\cite{Anders_2022}. For example, comparing the QNZL magnetization curves of Ni and Gd in Fig.~\ref{fig:equilibrium}(a) and Fig.~\ref{fig:equilibrium}(b), respectively, we see that for a large magnetic moment (Gd) the equilibration curve tends towards to classical ASD curve (orange line). 
In contrast, for a short spin length $S_0$ (Ni) the QNZL magnetization curve shows a clear difference to the classical ASD curve. While the low-temperature behavior of $m(T)$ for Fe and Co is well-reproduced by the QNZL noise, there are deviations at higher temperatures~Fig.\ref{fig:equilibrium}(c) and (d). 
For cobalt this is mainly attributed to the phase transition from fcc to hcp cobalt at $700$~K~\cite{Lizarraga2017}. 
This transition is not taken into account in the simulations of Fig.~\ref{fig:equilibrium}(d) as the full temperature range is simulated by a fixed parameter set detailed in Tab.~\ref{tab:parameters}.
Next to that, the deviation is likely caused by the larger moments of $\mu_{\mathrm{Co}} = 1.72 \mu_\mathrm{B}$ and $\mu_{\mathrm{Fe}} = 2.22 \mu_\mathrm{B}$ giving a closer correspondence to the classical m(T) curve that is more apparent at higher temperatures~\cite{Cerisola_2024}.

\section{Discussion and Outlook}
\label{sec:discussion}

In this work, we have implemented the open-system Landau-Lifshitz-Gilbert equation with colored quantum noise in the atomistic spin dynamics software package \textsc{vampire}, enabling the investigation of large-scale, exchange-coupled spin systems. We have implemented three distinct thermostats with Lorentzian power-spectral densities characterizing quantum, quantum without zero-point motion, and classical behaviors. In the classical limit, we reproduce the usual classical behavior of standard atomistic spin models. While the quantum thermostat correctly captures the temperature scaling of magnetization, it introduces zero-point fluctuations leading to non-unity magnetization at $T = 0$ K~\cite{Anders_2022}. The QNZL noise, which omits these fluctuations, provides the most accurate description of the temperature-dependent equilibrium magnetization for the studied materials at low temperatures, and is especially accurate for gadolinium and nickel over the full temperature range. Unlike previous approaches such as temperature-rescaling~\cite{Evans_2015}, the quantum thermostat naturally introduces the quantum mechanical nature of the heat bath from first principles, directly parameterized from material properties such as the phononic density of states.

In contrast to classical ASD simulations, in our \mLLG framework the spin length plays an important role, leading to a characteristic transition from quantum to classical temperature-dependent equilibrium magnetization for larger spin lengths. Assuming $S_0 = \mu_i$ within the QNZL thermostat, the large local moments of Fe and Co lead to more classical-like behavior than for small moments like Ni where $\mu_i \sim \hbar / 2$, in contrast to experimental observations where the $\mathrm{m}(T)$ for Co and Fe is more quantum-like~\cite{Evans_2015}. This suggests that an appropriate value for $S_0$ for Fe and Co to reproduce experimental observations could be significantly lower than the full value of the local moment. Including LSF to reflect the itinerant character of these materials could also bring the simulated $\mathrm{m}(T)$ curves into closer agreement with experiment~\cite{EllisPRB2019}.

The choice of Lorentzian parameters has a modest effect on the shape of the $\mathrm{m}(T)$ curve, which is mostly determined by the thermostat and spin length, but can have important effects on the dynamic response. A non-flat bath spectral density departs from the paradigm of purely Markovian magnetization dynamics of atomistic spin models. The non-Markovianity arises by incorporating time-correlated fluctuations and a memory kernel in the equation of motion. This approach opens new avenues for experimental validation and for probing memory effects in magnetic systems. The ability to parameterize the bath directly from material properties—such as phonon densities of states—ensures a first-principles basis for both equilibrium and dynamic simulations.

In summary, our implementation of quantum colored noise in atomistic spin dynamics significantly improves the modeling of equilibrium magnetization in local moment systems, thereby offering a more accurate description of their thermodynamic properties and establishing a foundation for future exploration of their dynamic behavior. The integration of fermionic (electron) baths alongside the current bosonic (phonon) approach represents a promising direction for capturing the complete spectrum of magnetic behavior across different materials and timescales. This can include effects such as ultrafast demagnetization~\cite{BeaurepairePRL1996}, nutation~\cite{Mondal_2023,Neeraj_2020,De_2024}, and coherent interaction between phonons and the magnetization~\cite{Unikandanunni_2022,Hartmann_2025}.

Further work could include a dynamic thermostat capable of modeling non-equilibrium magnetization dynamics with a quantum thermostat and time-correlated noise.

\section*{Acknowledgments}
We would like to thank J. Barker, C. Hogg, F. Cerisola, M. Strungaru, R. W. Chantrell and M. Berritta for insightful discussions and A. von Reppert for a critical reading of the manuscript.
We thank R. Lizárraga for generously sharing the cobalt density of state data. FCW and MB acknowledge the DFG for financial support through
Project No. 328545488—TRR 227, Project A10.
FH and JA acknowledge support from DFG, grants 513075417 and 384846402.
JA acknowledges support from EPSRC, grant EP/M009165/1, and the Royal Society.

\section*{Author contributions}
FCW implemented the \mLLG into the \textsc{vampire} software package with assistance from RFLE. FCW performed the atomistic calculations and analyzed the data. FH drafted the manuscript and developed the framework for the model, based on earlier work by JA.  JA and RFLE supervised the project. All authors contributed to the final manuscript and presentation of results.

\section*{Competing interests}
The authors declare no competing interests.

\section*{Data availability}
The implemented numerical method is publicly available in \href{https://github.com/richard-evans/vampire/tree/quantum-thermostat}{\textsc{Vampire}}.

\bibliography{main.bib}

\begin{thebibliography}{52}%
\makeatletter
\providecommand \@ifxundefined [1]{%
 \@ifx{#1\undefined}
}%
\providecommand \@ifnum [1]{%
 \ifnum #1\expandafter \@firstoftwo
 \else \expandafter \@secondoftwo
 \fi
}%
\providecommand \@ifx [1]{%
 \ifx #1\expandafter \@firstoftwo
 \else \expandafter \@secondoftwo
 \fi
}%
\providecommand \natexlab [1]{#1}%
\providecommand \enquote  [1]{``#1''}%
\providecommand \bibnamefont  [1]{#1}%
\providecommand \bibfnamefont [1]{#1}%
\providecommand \citenamefont [1]{#1}%
\providecommand \href@noop [0]{\@secondoftwo}%
\providecommand \href [0]{\begingroup \@sanitize@url \@href}%
\providecommand \@href[1]{\@@startlink{#1}\@@href}%
\providecommand \@@href[1]{\endgroup#1\@@endlink}%
\providecommand \@sanitize@url [0]{\catcode `\\12\catcode `\$12\catcode
  `\&12\catcode `\#12\catcode `\^12\catcode `\_12\catcode `\%12\relax}%
\providecommand \@@startlink[1]{}%
\providecommand \@@endlink[0]{}%
\providecommand \url  [0]{\begingroup\@sanitize@url \@url }%
\providecommand \@url [1]{\endgroup\@href {#1}{\urlprefix }}%
\providecommand \urlprefix  [0]{URL }%
\providecommand \Eprint [0]{\href }%
\providecommand \doibase [0]{https://doi.org/}%
\providecommand \selectlanguage [0]{\@gobble}%
\providecommand \bibinfo  [0]{\@secondoftwo}%
\providecommand \bibfield  [0]{\@secondoftwo}%
\providecommand \translation [1]{[#1]}%
\providecommand \BibitemOpen [0]{}%
\providecommand \bibitemStop [0]{}%
\providecommand \bibitemNoStop [0]{.\EOS\space}%
\providecommand \EOS [0]{\spacefactor3000\relax}%
\providecommand \BibitemShut  [1]{\csname bibitem#1\endcsname}%
\let\auto@bib@innerbib\@empty
\bibitem [{\citenamefont {Mryasov}\ \emph {et~al.}(2005)\citenamefont
  {Mryasov}, \citenamefont {Nowak}, \citenamefont {Guslienko},\ and\
  \citenamefont {Chantrell}}]{Mryasov_2005}%
  \BibitemOpen
  \bibfield  {author} {\bibinfo {author} {\bibfnamefont {O.~N.}\ \bibnamefont
  {Mryasov}}, \bibinfo {author} {\bibfnamefont {U.}~\bibnamefont {Nowak}},
  \bibinfo {author} {\bibfnamefont {K.~Y.}\ \bibnamefont {Guslienko}},\ and\
  \bibinfo {author} {\bibfnamefont {R.~W.}\ \bibnamefont {Chantrell}},\
  }\bibfield  {title} {\bibinfo {title} {Temperature-dependent magnetic
  properties of fept: Effective spin hamiltonian model},\ }\href
  {https://dx.doi.org/10.1209/epl/i2004-10404-2} {\bibfield  {journal}
  {\bibinfo  {journal} {EPL}\ }\textbf {\bibinfo {volume} {69}},\ \bibinfo
  {pages} {805} (\bibinfo {year} {2005})}\BibitemShut {NoStop}%
\bibitem [{\citenamefont {Ostler}\ \emph {et~al.}(2011)\citenamefont {Ostler},
  \citenamefont {Evans}, \citenamefont {Chantrell}, \citenamefont {Atxitia},
  \citenamefont {Chubykalo-Fesenko}, \citenamefont {Radu}, \citenamefont
  {Abrudan}, \citenamefont {Radu}, \citenamefont {Tsukamoto}, \citenamefont
  {Itoh}, \citenamefont {Kirilyuk}, \citenamefont {Rasing},\ and\ \citenamefont
  {Kimel}}]{Radu_2011}%
  \BibitemOpen
  \bibfield  {author} {\bibinfo {author} {\bibfnamefont {T.~A.}\ \bibnamefont
  {Ostler}}, \bibinfo {author} {\bibfnamefont {R.~F.~L.}\ \bibnamefont
  {Evans}}, \bibinfo {author} {\bibfnamefont {R.~W.}\ \bibnamefont
  {Chantrell}}, \bibinfo {author} {\bibfnamefont {U.}~\bibnamefont {Atxitia}},
  \bibinfo {author} {\bibfnamefont {O.}~\bibnamefont {Chubykalo-Fesenko}},
  \bibinfo {author} {\bibfnamefont {I.}~\bibnamefont {Radu}}, \bibinfo {author}
  {\bibfnamefont {R.}~\bibnamefont {Abrudan}}, \bibinfo {author} {\bibfnamefont
  {F.}~\bibnamefont {Radu}}, \bibinfo {author} {\bibfnamefont {A.}~\bibnamefont
  {Tsukamoto}}, \bibinfo {author} {\bibfnamefont {A.}~\bibnamefont {Itoh}},
  \bibinfo {author} {\bibfnamefont {A.}~\bibnamefont {Kirilyuk}}, \bibinfo
  {author} {\bibfnamefont {T.}~\bibnamefont {Rasing}},\ and\ \bibinfo {author}
  {\bibfnamefont {A.}~\bibnamefont {Kimel}},\ }\bibfield  {title} {\bibinfo
  {title} {Crystallographically amorphous ferrimagnetic alloys: Comparing a
  localized atomistic spin model with experiments},\ }\href
  {https://doi.org/10.1103/PhysRevB.84.024407} {\bibfield  {journal} {\bibinfo
  {journal} {Phys. Rev. B}\ }\textbf {\bibinfo {volume} {84}},\ \bibinfo
  {pages} {024407} (\bibinfo {year} {2011})}\BibitemShut {NoStop}%
\bibitem [{\citenamefont {Nedelkoski}\ \emph {et~al.}(2017)\citenamefont
  {Nedelkoski}, \citenamefont {Kepaptsoglou}, \citenamefont {Lari},
  \citenamefont {Wen}, \citenamefont {Booth}, \citenamefont {Oberdick},
  \citenamefont {Galindo}, \citenamefont {Ramasse}, \citenamefont {Evans},
  \citenamefont {Majetich},\ and\ \citenamefont {Lazarov}}]{Nedelkoski_2017}%
  \BibitemOpen
  \bibfield  {author} {\bibinfo {author} {\bibfnamefont {Z.}~\bibnamefont
  {Nedelkoski}}, \bibinfo {author} {\bibfnamefont {D.}~\bibnamefont
  {Kepaptsoglou}}, \bibinfo {author} {\bibfnamefont {L.}~\bibnamefont {Lari}},
  \bibinfo {author} {\bibfnamefont {T.}~\bibnamefont {Wen}}, \bibinfo {author}
  {\bibfnamefont {R.~A.}\ \bibnamefont {Booth}}, \bibinfo {author}
  {\bibfnamefont {S.~D.}\ \bibnamefont {Oberdick}}, \bibinfo {author}
  {\bibfnamefont {P.~L.}\ \bibnamefont {Galindo}}, \bibinfo {author}
  {\bibfnamefont {Q.~M.}\ \bibnamefont {Ramasse}}, \bibinfo {author}
  {\bibfnamefont {R.~F.~L.}\ \bibnamefont {Evans}}, \bibinfo {author}
  {\bibfnamefont {S.}~\bibnamefont {Majetich}},\ and\ \bibinfo {author}
  {\bibfnamefont {V.~K.}\ \bibnamefont {Lazarov}},\ }\bibfield  {title}
  {\bibinfo {title} {Origin of reduced magnetization and domain formation in
  small magnetite nanoparticles},\ }\href {https://doi.org/10.1038/srep45997}
  {\bibfield  {journal} {\bibinfo  {journal} {Sci. Rep.}\ }\textbf {\bibinfo
  {volume} {7}},\ \bibinfo {pages} {45997} (\bibinfo {year}
  {2017})}\BibitemShut {NoStop}%
\bibitem [{\citenamefont {Jenkins}\ \emph {et~al.}(2021)\citenamefont
  {Jenkins}, \citenamefont {Chantrell},\ and\ \citenamefont
  {Evans}}]{JenkinsPRM2021}%
  \BibitemOpen
  \bibfield  {author} {\bibinfo {author} {\bibfnamefont {S.}~\bibnamefont
  {Jenkins}}, \bibinfo {author} {\bibfnamefont {R.~W.}\ \bibnamefont
  {Chantrell}},\ and\ \bibinfo {author} {\bibfnamefont {R.~F.~L.}\ \bibnamefont
  {Evans}},\ }\bibfield  {title} {\bibinfo {title} {Atomistic simulations of
  the magnetic properties of
  ${\mathrm{ir}}_{x}{\mathrm{mn}}_{1\ensuremath{-}x}$ alloys},\ }\href
  {https://doi.org/10.1103/PhysRevMaterials.5.034406} {\bibfield  {journal}
  {\bibinfo  {journal} {Phys. Rev. Mater.}\ }\textbf {\bibinfo {volume} {5}},\
  \bibinfo {pages} {034406} (\bibinfo {year} {2021})}\BibitemShut {NoStop}%
\bibitem [{\citenamefont {Barker}\ and\ \citenamefont
  {Chantrell}(2015)}]{Barker_2015}%
  \BibitemOpen
  \bibfield  {author} {\bibinfo {author} {\bibfnamefont {J.}~\bibnamefont
  {Barker}}\ and\ \bibinfo {author} {\bibfnamefont {R.~W.}\ \bibnamefont
  {Chantrell}},\ }\bibfield  {title} {\bibinfo {title} {Higher-order exchange
  interactions leading to metamagnetism in ferh},\ }\href
  {https://link.aps.org/doi/10.1103/PhysRevB.92.094402} {\bibfield  {journal}
  {\bibinfo  {journal} {Phys. Rev. B}\ }\textbf {\bibinfo {volume} {92}},\
  \bibinfo {pages} {094402} (\bibinfo {year} {2015})}\BibitemShut {NoStop}%
\bibitem [{\citenamefont {Barker}\ and\ \citenamefont
  {Bauer}(2016)}]{Barker_2016}%
  \BibitemOpen
  \bibfield  {author} {\bibinfo {author} {\bibfnamefont {J.}~\bibnamefont
  {Barker}}\ and\ \bibinfo {author} {\bibfnamefont {G.~E.~W.}\ \bibnamefont
  {Bauer}},\ }\bibfield  {title} {\bibinfo {title} {Thermal spin dynamics of
  yttrium iron garnet},\ }\href
  {https://doi.org/10.1103/PhysRevLett.117.217201} {\bibfield  {journal}
  {\bibinfo  {journal} {Phys. Rev. Lett.}\ }\textbf {\bibinfo {volume} {117}},\
  \bibinfo {pages} {217201} (\bibinfo {year} {2016})}\BibitemShut {NoStop}%
\bibitem [{\citenamefont {Toga}\ \emph {et~al.}(2016)\citenamefont {Toga},
  \citenamefont {Matsumoto}, \citenamefont {Miyashita}, \citenamefont {Akai},
  \citenamefont {Doi}, \citenamefont {Miyake},\ and\ \citenamefont
  {Sakuma}}]{Toga_2016}%
  \BibitemOpen
  \bibfield  {author} {\bibinfo {author} {\bibfnamefont {Y.}~\bibnamefont
  {Toga}}, \bibinfo {author} {\bibfnamefont {M.}~\bibnamefont {Matsumoto}},
  \bibinfo {author} {\bibfnamefont {S.}~\bibnamefont {Miyashita}}, \bibinfo
  {author} {\bibfnamefont {H.}~\bibnamefont {Akai}}, \bibinfo {author}
  {\bibfnamefont {S.}~\bibnamefont {Doi}}, \bibinfo {author} {\bibfnamefont
  {T.}~\bibnamefont {Miyake}},\ and\ \bibinfo {author} {\bibfnamefont
  {A.}~\bibnamefont {Sakuma}},\ }\bibfield  {title} {\bibinfo {title} {Monte
  carlo analysis for finite-temperature magnetism of
  ${\mathrm{{n}d}}_{2}{\mathrm{{f}e}}_{14}\mathrm{B}$ permanent magnet},\
  }\href {https://link.aps.org/doi/10.1103/PhysRevB.94.174433} {\bibfield
  {journal} {\bibinfo  {journal} {Phys. Rev. B}\ }\textbf {\bibinfo {volume}
  {94}},\ \bibinfo {pages} {174433} (\bibinfo {year} {2016})}\BibitemShut
  {NoStop}%
\bibitem [{\citenamefont {Miyashita}\ \emph {et~al.}(2018)\citenamefont
  {Miyashita}, \citenamefont {Nishino}, \citenamefont {Toga}, \citenamefont
  {Hinokihara}, \citenamefont {Miyake}, \citenamefont {Hirosawa},\ and\
  \citenamefont {Sakuma}}]{Miyashita_2018}%
  \BibitemOpen
  \bibfield  {author} {\bibinfo {author} {\bibfnamefont {S.}~\bibnamefont
  {Miyashita}}, \bibinfo {author} {\bibfnamefont {M.}~\bibnamefont {Nishino}},
  \bibinfo {author} {\bibfnamefont {Y.}~\bibnamefont {Toga}}, \bibinfo {author}
  {\bibfnamefont {T.}~\bibnamefont {Hinokihara}}, \bibinfo {author}
  {\bibfnamefont {T.}~\bibnamefont {Miyake}}, \bibinfo {author} {\bibfnamefont
  {S.}~\bibnamefont {Hirosawa}},\ and\ \bibinfo {author} {\bibfnamefont
  {A.}~\bibnamefont {Sakuma}},\ }\bibfield  {title} {\bibinfo {title}
  {Perspectives of stochastic micromagnetism of {N}d2{F}e14b and computation of
  thermally activated reversal process},\ }\href
  {https://doi.org/https://doi.org/10.1016/j.scriptamat.2017.11.012} {\bibfield
   {journal} {\bibinfo  {journal} {Scr. Mater.}\ }\textbf {\bibinfo {volume}
  {154}},\ \bibinfo {pages} {259} (\bibinfo {year} {2018})}\BibitemShut
  {NoStop}%
\bibitem [{\citenamefont {Sampan-a pai}\ \emph {et~al.}(2019)\citenamefont
  {Sampan-a pai}, \citenamefont {Chureemart}, \citenamefont {Chantrell},
  \citenamefont {Chepulskyy}, \citenamefont {Wang}, \citenamefont {Apalkov},
  \citenamefont {Evans},\ and\ \citenamefont
  {Chureemart}}]{Sampan-a-paiPRA2019}%
  \BibitemOpen
  \bibfield  {author} {\bibinfo {author} {\bibfnamefont {S.}~\bibnamefont
  {Sampan-a pai}}, \bibinfo {author} {\bibfnamefont {J.}~\bibnamefont
  {Chureemart}}, \bibinfo {author} {\bibfnamefont {R.~W.}\ \bibnamefont
  {Chantrell}}, \bibinfo {author} {\bibfnamefont {R.}~\bibnamefont
  {Chepulskyy}}, \bibinfo {author} {\bibfnamefont {S.}~\bibnamefont {Wang}},
  \bibinfo {author} {\bibfnamefont {D.}~\bibnamefont {Apalkov}}, \bibinfo
  {author} {\bibfnamefont {R.~F.~L.}\ \bibnamefont {Evans}},\ and\ \bibinfo
  {author} {\bibfnamefont {P.}~\bibnamefont {Chureemart}},\ }\bibfield  {title}
  {\bibinfo {title} {Temperature and thickness dependence of statistical
  fluctuations of the gilbert damping in
  $\mathrm{Co}$-$\mathrm{Fe}$-$\mathrm{B}$/$\mathrm{Mg}\mathrm{O}$ bilayers},\
  }\href {https://doi.org/10.1103/PhysRevApplied.11.044001} {\bibfield
  {journal} {\bibinfo  {journal} {Phys. Rev. Appl.}\ }\textbf {\bibinfo
  {volume} {11}},\ \bibinfo {pages} {044001} (\bibinfo {year}
  {2019})}\BibitemShut {NoStop}%
\bibitem [{\citenamefont {Evans}\ \emph {et~al.}(2015)\citenamefont {Evans},
  \citenamefont {Atxitia},\ and\ \citenamefont {Chantrell}}]{Evans_2015}%
  \BibitemOpen
  \bibfield  {author} {\bibinfo {author} {\bibfnamefont {R.~F.~L.}\
  \bibnamefont {Evans}}, \bibinfo {author} {\bibfnamefont {U.}~\bibnamefont
  {Atxitia}},\ and\ \bibinfo {author} {\bibfnamefont {R.~W.}\ \bibnamefont
  {Chantrell}},\ }\bibfield  {title} {\bibinfo {title} {Quantitative simulation
  of temperature-dependent magnetization dynamics and equilibrium properties of
  elemental ferromagnets},\ }\href
  {http://dx.doi.org/10.1103/physrevb.91.144425} {\bibfield  {journal}
  {\bibinfo  {journal} {Phys. Rev. B}\ }\textbf {\bibinfo {volume} {91}},\
  \bibinfo {pages} {144425} (\bibinfo {year} {2015})}\BibitemShut {NoStop}%
\bibitem [{\citenamefont {Strungaru}\ \emph {et~al.}(2021)\citenamefont
  {Strungaru}, \citenamefont {Ellis}, \citenamefont {Ruta}, \citenamefont
  {Chubykalo-Fesenko}, \citenamefont {Evans},\ and\ \citenamefont
  {Chantrell}}]{Strungaru_2021}%
  \BibitemOpen
  \bibfield  {author} {\bibinfo {author} {\bibfnamefont {M.}~\bibnamefont
  {Strungaru}}, \bibinfo {author} {\bibfnamefont {M.~O.~A.}\ \bibnamefont
  {Ellis}}, \bibinfo {author} {\bibfnamefont {S.}~\bibnamefont {Ruta}},
  \bibinfo {author} {\bibfnamefont {O.}~\bibnamefont {Chubykalo-Fesenko}},
  \bibinfo {author} {\bibfnamefont {R.~F.~L.}\ \bibnamefont {Evans}},\ and\
  \bibinfo {author} {\bibfnamefont {R.~W.}\ \bibnamefont {Chantrell}},\
  }\bibfield  {title} {\bibinfo {title} {Spin-lattice dynamics model with
  angular momentum transfer for canonical and microcanonical ensembles},\
  }\href {https://doi.org/10.1103/PhysRevB.103.024429} {\bibfield  {journal}
  {\bibinfo  {journal} {Phys. Rev. B}\ }\textbf {\bibinfo {volume} {103}},\
  \bibinfo {pages} {024429} (\bibinfo {year} {2021})}\BibitemShut {NoStop}%
\bibitem [{\citenamefont {Ma}\ \emph {et~al.}(2008)\citenamefont {Ma},
  \citenamefont {Woo},\ and\ \citenamefont {Dudarev}}]{Ma_2008}%
  \BibitemOpen
  \bibfield  {author} {\bibinfo {author} {\bibfnamefont {P.-W.}\ \bibnamefont
  {Ma}}, \bibinfo {author} {\bibfnamefont {C.~H.}\ \bibnamefont {Woo}},\ and\
  \bibinfo {author} {\bibfnamefont {S.~L.}\ \bibnamefont {Dudarev}},\
  }\bibfield  {title} {\bibinfo {title} {Large-scale simulation of the
  spin-lattice dynamics in ferromagnetic iron},\ }\href
  {https://doi.org/10.1103/PhysRevB.78.024434} {\bibfield  {journal} {\bibinfo
  {journal} {Phys. Rev. B}\ }\textbf {\bibinfo {volume} {78}},\ \bibinfo
  {pages} {024434} (\bibinfo {year} {2008})}\BibitemShut {NoStop}%
\bibitem [{\citenamefont {Tranchida}\ \emph {et~al.}(2018)\citenamefont
  {Tranchida}, \citenamefont {Plimpton}, \citenamefont {Thibaudeau},\ and\
  \citenamefont {Thompson}}]{Tranchida_2018}%
  \BibitemOpen
  \bibfield  {author} {\bibinfo {author} {\bibfnamefont {J.}~\bibnamefont
  {Tranchida}}, \bibinfo {author} {\bibfnamefont {S.}~\bibnamefont {Plimpton}},
  \bibinfo {author} {\bibfnamefont {P.}~\bibnamefont {Thibaudeau}},\ and\
  \bibinfo {author} {\bibfnamefont {A.}~\bibnamefont {Thompson}},\ }\bibfield
  {title} {\bibinfo {title} {Massively parallel symplectic algorithm for
  coupled magnetic spin dynamics and molecular dynamics},\ }\href
  {http://dx.doi.org/10.1016/j.jcp.2018.06.042} {\bibfield  {journal} {\bibinfo
   {journal} {J. Comput. Phys.}\ }\textbf {\bibinfo {volume} {372}},\ \bibinfo
  {pages} {406–425} (\bibinfo {year} {2018})}\BibitemShut {NoStop}%
\bibitem [{\citenamefont {M\"uller}\ \emph {et~al.}(2019)\citenamefont
  {M\"uller}, \citenamefont {Hoffmann}, \citenamefont {Di\ss{}elkamp},
  \citenamefont {Sch\"urhoff}, \citenamefont {Mavros}, \citenamefont
  {Sallermann}, \citenamefont {Kiselev}, \citenamefont {J\'onsson},\ and\
  \citenamefont {Bl\"ugel}}]{Muller_2019}%
  \BibitemOpen
  \bibfield  {author} {\bibinfo {author} {\bibfnamefont {G.~P.}\ \bibnamefont
  {M\"uller}}, \bibinfo {author} {\bibfnamefont {M.}~\bibnamefont {Hoffmann}},
  \bibinfo {author} {\bibfnamefont {C.}~\bibnamefont {Di\ss{}elkamp}}, \bibinfo
  {author} {\bibfnamefont {D.}~\bibnamefont {Sch\"urhoff}}, \bibinfo {author}
  {\bibfnamefont {S.}~\bibnamefont {Mavros}}, \bibinfo {author} {\bibfnamefont
  {M.}~\bibnamefont {Sallermann}}, \bibinfo {author} {\bibfnamefont {N.~S.}\
  \bibnamefont {Kiselev}}, \bibinfo {author} {\bibfnamefont {H.}~\bibnamefont
  {J\'onsson}},\ and\ \bibinfo {author} {\bibfnamefont {S.}~\bibnamefont
  {Bl\"ugel}},\ }\bibfield  {title} {\bibinfo {title} {Spirit: Multifunctional
  framework for atomistic spin simulations},\ }\href
  {https://doi.org/10.1103/PhysRevB.99.224414} {\bibfield  {journal} {\bibinfo
  {journal} {Phys. Rev. B}\ }\textbf {\bibinfo {volume} {99}},\ \bibinfo
  {pages} {224414} (\bibinfo {year} {2019})}\BibitemShut {NoStop}%
\bibitem [{\citenamefont {Hellsvik}\ \emph {et~al.}(2019)\citenamefont
  {Hellsvik}, \citenamefont {Thonig}, \citenamefont {Modin}, \citenamefont
  {Iu\ifmmode~\mbox{\c{s}}\else \c{s}\fi{}an}, \citenamefont {Bergman},
  \citenamefont {Eriksson}, \citenamefont {Bergqvist},\ and\ \citenamefont
  {Delin}}]{Hellsvik_2019}%
  \BibitemOpen
  \bibfield  {author} {\bibinfo {author} {\bibfnamefont {J.}~\bibnamefont
  {Hellsvik}}, \bibinfo {author} {\bibfnamefont {D.}~\bibnamefont {Thonig}},
  \bibinfo {author} {\bibfnamefont {K.}~\bibnamefont {Modin}}, \bibinfo
  {author} {\bibfnamefont {D.}~\bibnamefont {Iu\ifmmode~\mbox{\c{s}}\else
  \c{s}\fi{}an}}, \bibinfo {author} {\bibfnamefont {A.}~\bibnamefont
  {Bergman}}, \bibinfo {author} {\bibfnamefont {O.}~\bibnamefont {Eriksson}},
  \bibinfo {author} {\bibfnamefont {L.}~\bibnamefont {Bergqvist}},\ and\
  \bibinfo {author} {\bibfnamefont {A.}~\bibnamefont {Delin}},\ }\bibfield
  {title} {\bibinfo {title} {General method for atomistic spin-lattice dynamics
  with first-principles accuracy},\ }\href
  {https://doi.org/10.1103/PhysRevB.99.104302} {\bibfield  {journal} {\bibinfo
  {journal} {Phys. Rev. B}\ }\textbf {\bibinfo {volume} {99}},\ \bibinfo
  {pages} {104302} (\bibinfo {year} {2019})}\BibitemShut {NoStop}%
\bibitem [{\citenamefont {Lange}\ \emph {et~al.}(2023)\citenamefont {Lange},
  \citenamefont {Mankovsky}, \citenamefont {Polesya}, \citenamefont
  {Wei\ss{}enhofer}, \citenamefont {Nowak},\ and\ \citenamefont
  {Ebert}}]{Lange_2023}%
  \BibitemOpen
  \bibfield  {author} {\bibinfo {author} {\bibfnamefont {H.}~\bibnamefont
  {Lange}}, \bibinfo {author} {\bibfnamefont {S.}~\bibnamefont {Mankovsky}},
  \bibinfo {author} {\bibfnamefont {S.}~\bibnamefont {Polesya}}, \bibinfo
  {author} {\bibfnamefont {M.}~\bibnamefont {Wei\ss{}enhofer}}, \bibinfo
  {author} {\bibfnamefont {U.}~\bibnamefont {Nowak}},\ and\ \bibinfo {author}
  {\bibfnamefont {H.}~\bibnamefont {Ebert}},\ }\bibfield  {title} {\bibinfo
  {title} {Calculating spin-lattice interactions in ferro- and
  antiferromagnets: The role of symmetry, dimension, and frustration},\ }\href
  {https://link.aps.org/doi/10.1103/PhysRevB.107.115176} {\bibfield  {journal}
  {\bibinfo  {journal} {Phys. Rev. B}\ }\textbf {\bibinfo {volume} {107}},\
  \bibinfo {pages} {115176} (\bibinfo {year} {2023})}\BibitemShut {NoStop}%
\bibitem [{\citenamefont {Ma}\ and\ \citenamefont {Dudarev}(2012)}]{Ma_2012}%
  \BibitemOpen
  \bibfield  {author} {\bibinfo {author} {\bibfnamefont {P.-W.}\ \bibnamefont
  {Ma}}\ and\ \bibinfo {author} {\bibfnamefont {S.~L.}\ \bibnamefont
  {Dudarev}},\ }\bibfield  {title} {\bibinfo {title} {Longitudinal magnetic
  fluctuations in langevin spin dynamics},\ }\href
  {http://dx.doi.org/10.1103/physrevb.86.054416} {\bibfield  {journal}
  {\bibinfo  {journal} {Phys. Rev. B}\ }\textbf {\bibinfo {volume} {86}},\
  \bibinfo {pages} {054416} (\bibinfo {year} {2012})}\BibitemShut {NoStop}%
\bibitem [{\citenamefont {Ellis}\ \emph {et~al.}(2019)\citenamefont {Ellis},
  \citenamefont {Galante},\ and\ \citenamefont {Sanvito}}]{EllisPRB2019}%
  \BibitemOpen
  \bibfield  {author} {\bibinfo {author} {\bibfnamefont {M.~O.~A.}\
  \bibnamefont {Ellis}}, \bibinfo {author} {\bibfnamefont {M.}~\bibnamefont
  {Galante}},\ and\ \bibinfo {author} {\bibfnamefont {S.}~\bibnamefont
  {Sanvito}},\ }\bibfield  {title} {\bibinfo {title} {Role of longitudinal
  fluctuations in $\mathrm{L}{1}_{0}$ fept},\ }\href
  {https://doi.org/10.1103/PhysRevB.100.214434} {\bibfield  {journal} {\bibinfo
   {journal} {Phys. Rev. B}\ }\textbf {\bibinfo {volume} {100}},\ \bibinfo
  {pages} {214434} (\bibinfo {year} {2019})}\BibitemShut {NoStop}%
\bibitem [{\citenamefont {Zahn}\ \emph {et~al.}(2021)\citenamefont {Zahn},
  \citenamefont {Jakobs}, \citenamefont {Windsor}, \citenamefont {Seiler},
  \citenamefont {Vasileiadis}, \citenamefont {Butcher}, \citenamefont {Qi},
  \citenamefont {Engel}, \citenamefont {Atxitia}, \citenamefont {Vorberger},\
  and\ \citenamefont {Ernstorfer}}]{Zahn_2021}%
  \BibitemOpen
  \bibfield  {author} {\bibinfo {author} {\bibfnamefont {D.}~\bibnamefont
  {Zahn}}, \bibinfo {author} {\bibfnamefont {F.}~\bibnamefont {Jakobs}},
  \bibinfo {author} {\bibfnamefont {Y.~W.}\ \bibnamefont {Windsor}}, \bibinfo
  {author} {\bibfnamefont {H.}~\bibnamefont {Seiler}}, \bibinfo {author}
  {\bibfnamefont {T.}~\bibnamefont {Vasileiadis}}, \bibinfo {author}
  {\bibfnamefont {T.~A.}\ \bibnamefont {Butcher}}, \bibinfo {author}
  {\bibfnamefont {Y.}~\bibnamefont {Qi}}, \bibinfo {author} {\bibfnamefont
  {D.}~\bibnamefont {Engel}}, \bibinfo {author} {\bibfnamefont
  {U.}~\bibnamefont {Atxitia}}, \bibinfo {author} {\bibfnamefont
  {J.}~\bibnamefont {Vorberger}},\ and\ \bibinfo {author} {\bibfnamefont
  {R.}~\bibnamefont {Ernstorfer}},\ }\bibfield  {title} {\bibinfo {title}
  {Lattice dynamics and ultrafast energy flow between electrons, spins, and
  phonons in a 3d ferromagnet},\ }\href
  {https://doi.org/10.1103/PhysRevResearch.3.023032} {\bibfield  {journal}
  {\bibinfo  {journal} {Phys. Rev. Res.}\ }\textbf {\bibinfo {volume} {3}},\
  \bibinfo {pages} {023032} (\bibinfo {year} {2021})}\BibitemShut {NoStop}%
\bibitem [{\citenamefont {Pankratova}\ \emph {et~al.}(2022)\citenamefont
  {Pankratova}, \citenamefont {Miranda}, \citenamefont {Thonig}, \citenamefont
  {Pereiro}, \citenamefont {Sj\"oqvist}, \citenamefont {Delin}, \citenamefont
  {Eriksson},\ and\ \citenamefont {Bergman}}]{Pankratova_2022}%
  \BibitemOpen
  \bibfield  {author} {\bibinfo {author} {\bibfnamefont {M.}~\bibnamefont
  {Pankratova}}, \bibinfo {author} {\bibfnamefont {I.~P.}\ \bibnamefont
  {Miranda}}, \bibinfo {author} {\bibfnamefont {D.}~\bibnamefont {Thonig}},
  \bibinfo {author} {\bibfnamefont {M.}~\bibnamefont {Pereiro}}, \bibinfo
  {author} {\bibfnamefont {E.}~\bibnamefont {Sj\"oqvist}}, \bibinfo {author}
  {\bibfnamefont {A.}~\bibnamefont {Delin}}, \bibinfo {author} {\bibfnamefont
  {O.}~\bibnamefont {Eriksson}},\ and\ \bibinfo {author} {\bibfnamefont
  {A.}~\bibnamefont {Bergman}},\ }\bibfield  {title} {\bibinfo {title}
  {Heat-conserving three-temperature model for ultrafast demagnetization in
  nickel},\ }\href {https://doi.org/10.1103/PhysRevB.106.174407} {\bibfield
  {journal} {\bibinfo  {journal} {Phys. Rev. B}\ }\textbf {\bibinfo {volume}
  {106}},\ \bibinfo {pages} {174407} (\bibinfo {year} {2022})}\BibitemShut
  {NoStop}%
\bibitem [{\citenamefont {Delczeg-Czirjak}\ \emph {et~al.}(2023)\citenamefont
  {Delczeg-Czirjak}, \citenamefont {Eriksson},\ and\ \citenamefont
  {Ruban}}]{Czirjak_2023}%
  \BibitemOpen
  \bibfield  {author} {\bibinfo {author} {\bibfnamefont {E.~K.}\ \bibnamefont
  {Delczeg-Czirjak}}, \bibinfo {author} {\bibfnamefont {O.}~\bibnamefont
  {Eriksson}},\ and\ \bibinfo {author} {\bibfnamefont {A.}~\bibnamefont
  {Ruban}},\ }\bibfield  {title} {\bibinfo {title} {The effect of longitudinal
  spin-fluctuations on high temperature properties of co3mn2ge},\ }\href
  {https://www.sciencedirect.com/science/article/pii/S1359646222007084}
  {\bibfield  {journal} {\bibinfo  {journal} {Scr. Mater.}\ }\textbf {\bibinfo
  {volume} {226}},\ \bibinfo {pages} {115213} (\bibinfo {year}
  {2023})}\BibitemShut {NoStop}%
\bibitem [{\citenamefont {Barker}\ and\ \citenamefont
  {Bauer}(2019)}]{Barker_2019}%
  \BibitemOpen
  \bibfield  {author} {\bibinfo {author} {\bibfnamefont {J.}~\bibnamefont
  {Barker}}\ and\ \bibinfo {author} {\bibfnamefont {G.~E.~W.}\ \bibnamefont
  {Bauer}},\ }\bibfield  {title} {\bibinfo {title} {Semiquantum thermodynamics
  of complex ferrimagnets},\ }\href
  {http://dx.doi.org/10.1103/physrevb.100.140401} {\bibfield  {journal}
  {\bibinfo  {journal} {Phys. Rev. B}\ }\textbf {\bibinfo {volume} {100}},\
  \bibinfo {pages} {140401} (\bibinfo {year} {2019})}\BibitemShut {NoStop}%
\bibitem [{\citenamefont {Bergqvist}\ and\ \citenamefont
  {Bergman}(2018)}]{BergqvistPRM2018}%
  \BibitemOpen
  \bibfield  {author} {\bibinfo {author} {\bibfnamefont {L.}~\bibnamefont
  {Bergqvist}}\ and\ \bibinfo {author} {\bibfnamefont {A.}~\bibnamefont
  {Bergman}},\ }\bibfield  {title} {\bibinfo {title} {Realistic finite
  temperature simulations of magnetic systems using quantum statistics},\
  }\href {https://doi.org/10.1103/PhysRevMaterials.2.013802} {\bibfield
  {journal} {\bibinfo  {journal} {Phys. Rev. Mater.}\ }\textbf {\bibinfo
  {volume} {2}},\ \bibinfo {pages} {013802} (\bibinfo {year}
  {2018})}\BibitemShut {NoStop}%
\bibitem [{\citenamefont {Berritta}\ \emph {et~al.}(2024)\citenamefont
  {Berritta}, \citenamefont {Scali}, \citenamefont {Cerisola},\ and\
  \citenamefont {Anders}}]{Berritta_2024}%
  \BibitemOpen
  \bibfield  {author} {\bibinfo {author} {\bibfnamefont {M.}~\bibnamefont
  {Berritta}}, \bibinfo {author} {\bibfnamefont {S.}~\bibnamefont {Scali}},
  \bibinfo {author} {\bibfnamefont {F.}~\bibnamefont {Cerisola}},\ and\
  \bibinfo {author} {\bibfnamefont {J.}~\bibnamefont {Anders}},\ }\bibfield
  {title} {\bibinfo {title} {Accounting for quantum effects in atomistic spin
  dynamics},\ }\href {http://dx.doi.org/10.1103/physrevb.109.174441} {\bibfield
   {journal} {\bibinfo  {journal} {Phys. Rev. B}\ }\textbf {\bibinfo {volume}
  {109}} (\bibinfo {year} {2024})}\BibitemShut {NoStop}%
\bibitem [{\citenamefont {Anders}\ \emph {et~al.}(2022)\citenamefont {Anders},
  \citenamefont {Sait},\ and\ \citenamefont {Horsley}}]{Anders_2022}%
  \BibitemOpen
  \bibfield  {author} {\bibinfo {author} {\bibfnamefont {J.}~\bibnamefont
  {Anders}}, \bibinfo {author} {\bibfnamefont {C.~R.~J.}\ \bibnamefont
  {Sait}},\ and\ \bibinfo {author} {\bibfnamefont {S.~A.~R.}\ \bibnamefont
  {Horsley}},\ }\bibfield  {title} {\bibinfo {title} {Quantum brownian motion
  for magnets},\ }\href {http://dx.doi.org/10.1088/1367-2630/ac4ef2} {\bibfield
   {journal} {\bibinfo  {journal} {New J. Phys.}\ }\textbf {\bibinfo {volume}
  {24}},\ \bibinfo {pages} {033020} (\bibinfo {year} {2022})}\BibitemShut
  {NoStop}%
\bibitem [{\citenamefont {Scali}\ \emph {et~al.}(2024)\citenamefont {Scali},
  \citenamefont {Horsley}, \citenamefont {Anders},\ and\ \citenamefont
  {Cerisola}}]{Scali_2024}%
  \BibitemOpen
  \bibfield  {author} {\bibinfo {author} {\bibfnamefont {S.}~\bibnamefont
  {Scali}}, \bibinfo {author} {\bibfnamefont {S.}~\bibnamefont {Horsley}},
  \bibinfo {author} {\bibfnamefont {J.}~\bibnamefont {Anders}},\ and\ \bibinfo
  {author} {\bibfnamefont {F.}~\bibnamefont {Cerisola}},\ }\bibfield  {title}
  {\bibinfo {title} {Spidy.jl: open-source julia package for the study of
  non-markovian stochastic dynamics},\ }\href
  {http://dx.doi.org/10.21105/joss.06263} {\bibfield  {journal} {\bibinfo
  {journal} {J. Open Source Softw.}\ }\textbf {\bibinfo {volume} {9}},\
  \bibinfo {pages} {6263} (\bibinfo {year} {2024})}\BibitemShut {NoStop}%
\bibitem [{\citenamefont {Evans}\ \emph {et~al.}(2014)\citenamefont {Evans},
  \citenamefont {Fan}, \citenamefont {Chureemart}, \citenamefont {Ostler},
  \citenamefont {Ellis},\ and\ \citenamefont {Chantrell}}]{Evans_2014}%
  \BibitemOpen
  \bibfield  {author} {\bibinfo {author} {\bibfnamefont {R.~F.~L.}\
  \bibnamefont {Evans}}, \bibinfo {author} {\bibfnamefont {W.~J.}\ \bibnamefont
  {Fan}}, \bibinfo {author} {\bibfnamefont {P.}~\bibnamefont {Chureemart}},
  \bibinfo {author} {\bibfnamefont {T.~A.}\ \bibnamefont {Ostler}}, \bibinfo
  {author} {\bibfnamefont {M.~O.~A.}\ \bibnamefont {Ellis}},\ and\ \bibinfo
  {author} {\bibfnamefont {R.~W.}\ \bibnamefont {Chantrell}},\ }\bibfield
  {title} {\bibinfo {title} {Atomistic spin model simulations of magnetic
  nanomaterials},\ }\href {http://dx.doi.org/10.1088/0953-8984/26/10/103202}
  {\bibfield  {journal} {\bibinfo  {journal} {J. Phys. Condens. Matter}\
  }\textbf {\bibinfo {volume} {26}},\ \bibinfo {pages} {103202} (\bibinfo
  {year} {2014})}\BibitemShut {NoStop}%
\bibitem [{\citenamefont {Bloch}(1930)}]{Bloch1930}%
  \BibitemOpen
  \bibfield  {author} {\bibinfo {author} {\bibfnamefont {F.}~\bibnamefont
  {Bloch}},\ }\bibfield  {title} {\bibinfo {title} {Zur $\mathrm{T}$heorie des
  $\mathrm{F}$erromagnetismus},\ }\href {https://doi.org/10.1007/bf01339661}
  {\bibfield  {journal} {\bibinfo  {journal} {Zeitschrift f{\"u}r Physik}\
  }\textbf {\bibinfo {volume} {61}},\ \bibinfo {pages} {206–219} (\bibinfo
  {year} {1930})}\BibitemShut {NoStop}%
\bibitem [{\citenamefont {Sun}\ \emph {et~al.}(2025)\citenamefont {Sun},
  \citenamefont {Kang}, \citenamefont {Nuomin}, \citenamefont {Schwartz},
  \citenamefont {Beratan}, \citenamefont {Brown},\ and\ \citenamefont
  {Kim}}]{Sun2025}%
  \BibitemOpen
  \bibfield  {author} {\bibinfo {author} {\bibfnamefont {K.}~\bibnamefont
  {Sun}}, \bibinfo {author} {\bibfnamefont {M.}~\bibnamefont {Kang}}, \bibinfo
  {author} {\bibfnamefont {H.}~\bibnamefont {Nuomin}}, \bibinfo {author}
  {\bibfnamefont {G.}~\bibnamefont {Schwartz}}, \bibinfo {author}
  {\bibfnamefont {D.~N.}\ \bibnamefont {Beratan}}, \bibinfo {author}
  {\bibfnamefont {K.~R.}\ \bibnamefont {Brown}},\ and\ \bibinfo {author}
  {\bibfnamefont {J.}~\bibnamefont {Kim}},\ }\bibfield  {title} {\bibinfo
  {title} {Quantum simulation of spin-boson models with structured bath},\
  }\href {http://dx.doi.org/10.1038/s41467-025-59296-y} {\bibfield  {journal}
  {\bibinfo  {journal} {Nat. Commun.}\ }\textbf {\bibinfo {volume} {16}}
  (\bibinfo {year} {2025})}\BibitemShut {NoStop}%
\bibitem [{\citenamefont {Chen}\ \emph {et~al.}(2023)\citenamefont {Chen},
  \citenamefont {Yan}, \citenamefont {Gelin},\ and\ \citenamefont
  {Lü}}]{Chen2023}%
  \BibitemOpen
  \bibfield  {author} {\bibinfo {author} {\bibfnamefont {L.}~\bibnamefont
  {Chen}}, \bibinfo {author} {\bibfnamefont {Y.}~\bibnamefont {Yan}}, \bibinfo
  {author} {\bibfnamefont {M.~F.}\ \bibnamefont {Gelin}},\ and\ \bibinfo
  {author} {\bibfnamefont {Z.}~\bibnamefont {Lü}},\ }\bibfield  {title}
  {\bibinfo {title} {Dynamics of the spin-boson model: The effect of bath
  initial conditions},\ }\href {http://dx.doi.org/10.1063/5.0138399} {\bibfield
   {journal} {\bibinfo  {journal} {J. Chem. Phys}\ }\textbf {\bibinfo {volume}
  {158}} (\bibinfo {year} {2023})}\BibitemShut {NoStop}%
\bibitem [{\citenamefont {Hogg}\ \emph {et~al.}(2024)\citenamefont {Hogg},
  \citenamefont {Cerisola}, \citenamefont {Cresser}, \citenamefont {Horsley},\
  and\ \citenamefont {Anders}}]{Hogg2024}%
  \BibitemOpen
  \bibfield  {author} {\bibinfo {author} {\bibfnamefont {C.~R.}\ \bibnamefont
  {Hogg}}, \bibinfo {author} {\bibfnamefont {F.}~\bibnamefont {Cerisola}},
  \bibinfo {author} {\bibfnamefont {J.~D.}\ \bibnamefont {Cresser}}, \bibinfo
  {author} {\bibfnamefont {S.~A.~R.}\ \bibnamefont {Horsley}},\ and\ \bibinfo
  {author} {\bibfnamefont {J.}~\bibnamefont {Anders}},\ }\bibfield  {title}
  {\bibinfo {title} {Enhanced entanglement in multi-bath spin-boson models},\
  }\href {http://dx.doi.org/10.22331/q-2024-05-23-1357} {\bibfield  {journal}
  {\bibinfo  {journal} {Quantum}\ }\textbf {\bibinfo {volume} {8}},\ \bibinfo
  {pages} {1357} (\bibinfo {year} {2024})}\BibitemShut {NoStop}%
\bibitem [{\citenamefont {Born}\ \emph {et~al.}(2021)\citenamefont {Born},
  \citenamefont {Decker}, \citenamefont {Büchner}, \citenamefont {Haverkamp},
  \citenamefont {Ruotsalainen}, \citenamefont {Bauer}, \citenamefont
  {Pietzsch},\ and\ \citenamefont {Föhlisch}}]{Born2021}%
  \BibitemOpen
  \bibfield  {author} {\bibinfo {author} {\bibfnamefont {A.}~\bibnamefont
  {Born}}, \bibinfo {author} {\bibfnamefont {R.}~\bibnamefont {Decker}},
  \bibinfo {author} {\bibfnamefont {R.}~\bibnamefont {Büchner}}, \bibinfo
  {author} {\bibfnamefont {R.}~\bibnamefont {Haverkamp}}, \bibinfo {author}
  {\bibfnamefont {K.}~\bibnamefont {Ruotsalainen}}, \bibinfo {author}
  {\bibfnamefont {K.}~\bibnamefont {Bauer}}, \bibinfo {author} {\bibfnamefont
  {A.}~\bibnamefont {Pietzsch}},\ and\ \bibinfo {author} {\bibfnamefont
  {A.}~\bibnamefont {Föhlisch}},\ }\bibfield  {title} {\bibinfo {title}
  {Thresholding of the elliott-yafet spin-flip scattering in multi-sublattice
  magnets by the respective exchange energies},\ }\href
  {http://dx.doi.org/10.1038/s41598-021-81177-9} {\bibfield  {journal}
  {\bibinfo  {journal} {Sci. Rep.}\ }\textbf {\bibinfo {volume} {11}} (\bibinfo
  {year} {2021})}\BibitemShut {NoStop}%
\bibitem [{\citenamefont {Nemati}\ \emph {et~al.}(2022)\citenamefont {Nemati},
  \citenamefont {Henkel},\ and\ \citenamefont {Anders}}]{Nemati_2022}%
  \BibitemOpen
  \bibfield  {author} {\bibinfo {author} {\bibfnamefont {S.}~\bibnamefont
  {Nemati}}, \bibinfo {author} {\bibfnamefont {C.}~\bibnamefont {Henkel}},\
  and\ \bibinfo {author} {\bibfnamefont {J.}~\bibnamefont {Anders}},\
  }\bibfield  {title} {\bibinfo {title} {Coupling function from bath density of
  states},\ }\href {http://dx.doi.org/10.1209/0295-5075/ac7b42} {\bibfield
  {journal} {\bibinfo  {journal} {EPL}\ }\textbf {\bibinfo {volume} {139}},\
  \bibinfo {pages} {36002} (\bibinfo {year} {2022})}\BibitemShut {NoStop}%
\bibitem [{\citenamefont {Pershin}\ and\ \citenamefont
  {Di~Ventra}(2011)}]{Pershin_2011}%
  \BibitemOpen
  \bibfield  {author} {\bibinfo {author} {\bibfnamefont {Y.~V.}\ \bibnamefont
  {Pershin}}\ and\ \bibinfo {author} {\bibfnamefont {M.}~\bibnamefont
  {Di~Ventra}},\ }\bibfield  {title} {\bibinfo {title} {Memory effects in
  complex materials and nanoscale systems},\ }\href
  {http://dx.doi.org/10.1080/00018732.2010.544961} {\bibfield  {journal}
  {\bibinfo  {journal} {Adv. Phys.}\ }\textbf {\bibinfo {volume} {60}},\
  \bibinfo {pages} {145–227} (\bibinfo {year} {2011})}\BibitemShut {NoStop}%
\bibitem [{\citenamefont {Atxitia}\ \emph {et~al.}(2009)\citenamefont
  {Atxitia}, \citenamefont {Chubykalo-Fesenko}, \citenamefont {Chantrell},
  \citenamefont {Nowak},\ and\ \citenamefont {Rebei}}]{Atxitia_2009}%
  \BibitemOpen
  \bibfield  {author} {\bibinfo {author} {\bibfnamefont {U.}~\bibnamefont
  {Atxitia}}, \bibinfo {author} {\bibfnamefont {O.}~\bibnamefont
  {Chubykalo-Fesenko}}, \bibinfo {author} {\bibfnamefont {R.~W.}\ \bibnamefont
  {Chantrell}}, \bibinfo {author} {\bibfnamefont {U.}~\bibnamefont {Nowak}},\
  and\ \bibinfo {author} {\bibfnamefont {A.}~\bibnamefont {Rebei}},\ }\bibfield
   {title} {\bibinfo {title} {Ultrafast spin dynamics: The effect of colored
  noise},\ }\href {http://dx.doi.org/10.1103/physrevlett.102.057203} {\bibfield
   {journal} {\bibinfo  {journal} {Phys. Rev. Lett.}\ }\textbf {\bibinfo
  {volume} {102}},\ \bibinfo {pages} {057203} (\bibinfo {year}
  {2009})}\BibitemShut {NoStop}%
\bibitem [{\citenamefont {Frigo}\ and\ \citenamefont
  {Johnson}(2005)}]{Frigo2005}%
  \BibitemOpen
  \bibfield  {author} {\bibinfo {author} {\bibfnamefont {M.}~\bibnamefont
  {Frigo}}\ and\ \bibinfo {author} {\bibfnamefont {S.}~\bibnamefont
  {Johnson}},\ }\bibfield  {title} {\bibinfo {title} {The design and
  implementation of fftw3},\ }\href {https://doi.org/10.1109/jproc.2004.840301}
  {\bibfield  {journal} {\bibinfo  {journal} {Proc. IEEE}\ }\textbf {\bibinfo
  {volume} {93}},\ \bibinfo {pages} {216–231} (\bibinfo {year}
  {2005})}\BibitemShut {NoStop}%
\bibitem [{\citenamefont {Callen}\ and\ \citenamefont
  {Welton}(1951)}]{Callen1951}%
  \BibitemOpen
  \bibfield  {author} {\bibinfo {author} {\bibfnamefont {H.~B.}\ \bibnamefont
  {Callen}}\ and\ \bibinfo {author} {\bibfnamefont {T.~A.}\ \bibnamefont
  {Welton}},\ }\bibfield  {title} {\bibinfo {title} {Irreversibility and
  generalized noise},\ }\href {https://doi.org/10.1103/PhysRev.83.34}
  {\bibfield  {journal} {\bibinfo  {journal} {Phys. Rev.}\ }\textbf {\bibinfo
  {volume} {83}},\ \bibinfo {pages} {34} (\bibinfo {year} {1951})}\BibitemShut
  {NoStop}%
\bibitem [{\citenamefont {Kresch}\ \emph {et~al.}(2007)\citenamefont {Kresch},
  \citenamefont {Delaire}, \citenamefont {Stevens}, \citenamefont {Lin},\ and\
  \citenamefont {Fultz}}]{Kresch_2007}%
  \BibitemOpen
  \bibfield  {author} {\bibinfo {author} {\bibfnamefont {M.}~\bibnamefont
  {Kresch}}, \bibinfo {author} {\bibfnamefont {O.}~\bibnamefont {Delaire}},
  \bibinfo {author} {\bibfnamefont {R.}~\bibnamefont {Stevens}}, \bibinfo
  {author} {\bibfnamefont {J.}~\bibnamefont {Lin}},\ and\ \bibinfo {author}
  {\bibfnamefont {B.}~\bibnamefont {Fultz}},\ }\bibfield  {title} {\bibinfo
  {title} {Neutron scattering measurements of phonons in nickel at elevated
  temperatures},\ }\href {https://doi.org/10.1103/PHYSREVB.75.104301}
  {\bibfield  {journal} {\bibinfo  {journal} {Phys. Rev. B}\ }\textbf {\bibinfo
  {volume} {75}} (\bibinfo {year} {2007})}\BibitemShut {NoStop}%
\bibitem [{\citenamefont {Kuz'min}(2005)}]{Kuz_2005}%
  \BibitemOpen
  \bibfield  {author} {\bibinfo {author} {\bibfnamefont {M.~D.}\ \bibnamefont
  {Kuz'min}},\ }\bibfield  {title} {\bibinfo {title} {Shape of temperature
  dependence of spontaneous magnetization of ferromagnets: Quantitative
  analysis},\ }\href {https://link.aps.org/doi/10.1103/PhysRevLett.94.107204}
  {\bibfield  {journal} {\bibinfo  {journal} {Phys. Rev. Lett.}\ }\textbf
  {\bibinfo {volume} {94}},\ \bibinfo {pages} {107204} (\bibinfo {year}
  {2005})}\BibitemShut {NoStop}%
\bibitem [{\citenamefont {Sato}\ \emph {et~al.}(2018)\citenamefont {Sato},
  \citenamefont {Chureemart}, \citenamefont {Matsukura}, \citenamefont
  {Chantrell}, \citenamefont {Ohno},\ and\ \citenamefont
  {Evans}}]{SatoPRB2018}%
  \BibitemOpen
  \bibfield  {author} {\bibinfo {author} {\bibfnamefont {H.}~\bibnamefont
  {Sato}}, \bibinfo {author} {\bibfnamefont {P.}~\bibnamefont {Chureemart}},
  \bibinfo {author} {\bibfnamefont {F.}~\bibnamefont {Matsukura}}, \bibinfo
  {author} {\bibfnamefont {R.~W.}\ \bibnamefont {Chantrell}}, \bibinfo {author}
  {\bibfnamefont {H.}~\bibnamefont {Ohno}},\ and\ \bibinfo {author}
  {\bibfnamefont {R.~F.~L.}\ \bibnamefont {Evans}},\ }\bibfield  {title}
  {\bibinfo {title} {Temperature-dependent properties of cofeb/mgo thin films:
  Experiments versus simulations},\ }\href
  {https://link.aps.org/doi/10.1103/PhysRevB.98.214428} {\bibfield  {journal}
  {\bibinfo  {journal} {Phys. Rev. B}\ }\textbf {\bibinfo {volume} {98}},\
  \bibinfo {pages} {214428} (\bibinfo {year} {2018})}\BibitemShut {NoStop}%
\bibitem [{\citenamefont {Alp}\ \emph {et~al.}(2002)\citenamefont {Alp},
  \citenamefont {Sturhahn}, \citenamefont {Toellner}, \citenamefont {Zhao},
  \citenamefont {Hu},\ and\ \citenamefont {Brown}}]{Alp_2002}%
  \BibitemOpen
  \bibfield  {author} {\bibinfo {author} {\bibfnamefont {E.}~\bibnamefont
  {Alp}}, \bibinfo {author} {\bibfnamefont {W.}~\bibnamefont {Sturhahn}},
  \bibinfo {author} {\bibfnamefont {T.}~\bibnamefont {Toellner}}, \bibinfo
  {author} {\bibfnamefont {J.}~\bibnamefont {Zhao}}, \bibinfo {author}
  {\bibfnamefont {M.}~\bibnamefont {Hu}},\ and\ \bibinfo {author}
  {\bibfnamefont {D.}~\bibnamefont {Brown}},\ }\bibfield  {title} {\bibinfo
  {title} {Vibrational dynamics studies by nuclear resonant inelastic x-ray
  scattering},\ }\href {https://doi.org/10.1023/A:1025452401501} {\bibfield
  {journal} {\bibinfo  {journal} {Hyperfine Interact.}\ }\textbf {\bibinfo
  {volume} {144/145}},\ \bibinfo {pages} {3} (\bibinfo {year}
  {2002})}\BibitemShut {NoStop}%
\bibitem [{\citenamefont {Liz{\'a}rraga}\ \emph {et~al.}(2007)\citenamefont
  {Liz{\'a}rraga}, \citenamefont {Pan}, \citenamefont {Bergqvist},
  \citenamefont {Holmstr{\"o}m}, \citenamefont {Gercsi},\ and\ \citenamefont
  {Vitos}}]{Lizarraga2017}%
  \BibitemOpen
  \bibfield  {author} {\bibinfo {author} {\bibfnamefont {R.}~\bibnamefont
  {Liz{\'a}rraga}}, \bibinfo {author} {\bibfnamefont {F.}~\bibnamefont {Pan}},
  \bibinfo {author} {\bibfnamefont {L.}~\bibnamefont {Bergqvist}}, \bibinfo
  {author} {\bibfnamefont {E.}~\bibnamefont {Holmstr{\"o}m}}, \bibinfo {author}
  {\bibfnamefont {Z.}~\bibnamefont {Gercsi}},\ and\ \bibinfo {author}
  {\bibfnamefont {L.}~\bibnamefont {Vitos}},\ }\bibfield  {title} {\bibinfo
  {title} {First principles theory of the hcp-fcc phase transition in cobalt},\
  }\href {https://doi.org/https://doi.org/10.1038/s41598-017-03877-5}
  {\bibfield  {journal} {\bibinfo  {journal} {Sci. Rep.}\ }\textbf {\bibinfo
  {volume} {7}},\ \bibinfo {pages} {3778} (\bibinfo {year} {2007})}\BibitemShut
  {NoStop}%
\bibitem [{\citenamefont {Li}\ \emph {et~al.}(2021)\citenamefont {Li},
  \citenamefont {Ehteshami}, \citenamefont {Munro}, \citenamefont {Marqu\'es},
  \citenamefont {McMahon}, \citenamefont {MacLeod},\ and\ \citenamefont
  {Ackland}}]{Li_2021}%
  \BibitemOpen
  \bibfield  {author} {\bibinfo {author} {\bibfnamefont {Q.}~\bibnamefont
  {Li}}, \bibinfo {author} {\bibfnamefont {H.}~\bibnamefont {Ehteshami}},
  \bibinfo {author} {\bibfnamefont {K.}~\bibnamefont {Munro}}, \bibinfo
  {author} {\bibfnamefont {M.}~\bibnamefont {Marqu\'es}}, \bibinfo {author}
  {\bibfnamefont {M.~I.}\ \bibnamefont {McMahon}}, \bibinfo {author}
  {\bibfnamefont {S.~G.}\ \bibnamefont {MacLeod}},\ and\ \bibinfo {author}
  {\bibfnamefont {G.~J.}\ \bibnamefont {Ackland}},\ }\bibfield  {title}
  {\bibinfo {title} {Nonexistence of the $s\text{\ensuremath{-}}f$
  volume-collapse transition in solid gadolinium at pressure},\ }\href
  {https://link.aps.org/doi/10.1103/PhysRevB.104.144108} {\bibfield  {journal}
  {\bibinfo  {journal} {Phys. Rev. B}\ }\textbf {\bibinfo {volume} {104}},\
  \bibinfo {pages} {144108} (\bibinfo {year} {2021})}\BibitemShut {NoStop}%
\bibitem [{\citenamefont {Crangle}\ \emph {et~al.}(1971)\citenamefont
  {Crangle}, \citenamefont {Goodman},\ and\ \citenamefont
  {Sucksmith}}]{Crangle_1971}%
  \BibitemOpen
  \bibfield  {author} {\bibinfo {author} {\bibfnamefont {J.}~\bibnamefont
  {Crangle}}, \bibinfo {author} {\bibfnamefont {G.~M.}\ \bibnamefont
  {Goodman}},\ and\ \bibinfo {author} {\bibfnamefont {W.}~\bibnamefont
  {Sucksmith}},\ }\bibfield  {title} {\bibinfo {title} {The magnetization of
  pure iron and nickel},\ }\href {https://doi.org/10.1098/rspa.1971.0044}
  {\bibfield  {journal} {\bibinfo  {journal} {Proc. R. Soc. Lond. A}\ }\textbf
  {\bibinfo {volume} {321}},\ \bibinfo {pages} {477} (\bibinfo {year}
  {1971})}\BibitemShut {NoStop}%
\bibitem [{\citenamefont {Nigh}\ \emph {et~al.}(1963)\citenamefont {Nigh},
  \citenamefont {Legvold},\ and\ \citenamefont {Spedding}}]{Nigh_1963}%
  \BibitemOpen
  \bibfield  {author} {\bibinfo {author} {\bibfnamefont {H.~E.}\ \bibnamefont
  {Nigh}}, \bibinfo {author} {\bibfnamefont {S.}~\bibnamefont {Legvold}},\ and\
  \bibinfo {author} {\bibfnamefont {F.~H.}\ \bibnamefont {Spedding}},\
  }\bibfield  {title} {\bibinfo {title} {Magnetization and electrical
  resistivity of gadolinium single crystals},\ }\href
  {https://link.aps.org/doi/10.1103/PhysRev.132.1092} {\bibfield  {journal}
  {\bibinfo  {journal} {Phys. Rev.}\ }\textbf {\bibinfo {volume} {132}},\
  \bibinfo {pages} {1092} (\bibinfo {year} {1963})}\BibitemShut {NoStop}%
\bibitem [{\citenamefont {Cerisola}\ \emph {et~al.}(2024)\citenamefont
  {Cerisola}, \citenamefont {Berritta}, \citenamefont {Scali}, \citenamefont
  {Horsley}, \citenamefont {Cresser},\ and\ \citenamefont
  {Anders}}]{Cerisola_2024}%
  \BibitemOpen
  \bibfield  {author} {\bibinfo {author} {\bibfnamefont {F.}~\bibnamefont
  {Cerisola}}, \bibinfo {author} {\bibfnamefont {M.}~\bibnamefont {Berritta}},
  \bibinfo {author} {\bibfnamefont {S.}~\bibnamefont {Scali}}, \bibinfo
  {author} {\bibfnamefont {S.~A.~R.}\ \bibnamefont {Horsley}}, \bibinfo
  {author} {\bibfnamefont {J.~D.}\ \bibnamefont {Cresser}},\ and\ \bibinfo
  {author} {\bibfnamefont {J.}~\bibnamefont {Anders}},\ }\bibfield  {title}
  {\bibinfo {title} {Quantum–classical correspondence in spin–boson
  equilibrium states at arbitrary coupling},\ }\href
  {http://dx.doi.org/10.1088/1367-2630/ad4818} {\bibfield  {journal} {\bibinfo
  {journal} {New J. Phys.}\ }\textbf {\bibinfo {volume} {26}},\ \bibinfo
  {pages} {053032} (\bibinfo {year} {2024})}\BibitemShut {NoStop}%
\bibitem [{\citenamefont {Beaurepaire}\ \emph {et~al.}(1996)\citenamefont
  {Beaurepaire}, \citenamefont {Merle}, \citenamefont {Daunois},\ and\
  \citenamefont {Bigot}}]{BeaurepairePRL1996}%
  \BibitemOpen
  \bibfield  {author} {\bibinfo {author} {\bibfnamefont {E.}~\bibnamefont
  {Beaurepaire}}, \bibinfo {author} {\bibfnamefont {J.-C.}\ \bibnamefont
  {Merle}}, \bibinfo {author} {\bibfnamefont {A.}~\bibnamefont {Daunois}},\
  and\ \bibinfo {author} {\bibfnamefont {J.-Y.}\ \bibnamefont {Bigot}},\
  }\bibfield  {title} {\bibinfo {title} {Ultrafast spin dynamics in
  ferromagnetic nickel},\ }\href
  {https://link.aps.org/doi/10.1103/PhysRevLett.76.4250} {\bibfield  {journal}
  {\bibinfo  {journal} {Phys. Rev. Lett.}\ }\textbf {\bibinfo {volume} {76}},\
  \bibinfo {pages} {4250} (\bibinfo {year} {1996})}\BibitemShut {NoStop}%
\bibitem [{\citenamefont {Mondal}\ \emph {et~al.}(2023)\citenamefont {Mondal},
  \citenamefont {Rózsa}, \citenamefont {Farle}, \citenamefont {Oppeneer},
  \citenamefont {Nowak},\ and\ \citenamefont {Cherkasskii}}]{Mondal_2023}%
  \BibitemOpen
  \bibfield  {author} {\bibinfo {author} {\bibfnamefont {R.}~\bibnamefont
  {Mondal}}, \bibinfo {author} {\bibfnamefont {L.}~\bibnamefont {Rózsa}},
  \bibinfo {author} {\bibfnamefont {M.}~\bibnamefont {Farle}}, \bibinfo
  {author} {\bibfnamefont {P.~M.}\ \bibnamefont {Oppeneer}}, \bibinfo {author}
  {\bibfnamefont {U.}~\bibnamefont {Nowak}},\ and\ \bibinfo {author}
  {\bibfnamefont {M.}~\bibnamefont {Cherkasskii}},\ }\bibfield  {title}
  {\bibinfo {title} {Inertial effects in ultrafast spin dynamics},\ }\href
  {http://dx.doi.org/10.1016/j.jmmm.2023.170830} {\bibfield  {journal}
  {\bibinfo  {journal} {J. Magn. Magn. Mat.}\ }\textbf {\bibinfo {volume}
  {579}},\ \bibinfo {pages} {170830} (\bibinfo {year} {2023})}\BibitemShut
  {NoStop}%
\bibitem [{\citenamefont {Neeraj}\ \emph {et~al.}(2020)\citenamefont {Neeraj},
  \citenamefont {Awari}, \citenamefont {Kovalev}, \citenamefont {Polley},
  \citenamefont {Zhou~Hagström}, \citenamefont {Arekapudi}, \citenamefont
  {Semisalova}, \citenamefont {Lenz}, \citenamefont {Green}, \citenamefont
  {Deinert}, \citenamefont {Ilyakov}, \citenamefont {Chen}, \citenamefont
  {Bawatna}, \citenamefont {Scalera}, \citenamefont {d’Aquino}, \citenamefont
  {Serpico}, \citenamefont {Hellwig}, \citenamefont {Wegrowe}, \citenamefont
  {Gensch},\ and\ \citenamefont {Bonetti}}]{Neeraj_2020}%
  \BibitemOpen
  \bibfield  {author} {\bibinfo {author} {\bibfnamefont {K.}~\bibnamefont
  {Neeraj}}, \bibinfo {author} {\bibfnamefont {N.}~\bibnamefont {Awari}},
  \bibinfo {author} {\bibfnamefont {S.}~\bibnamefont {Kovalev}}, \bibinfo
  {author} {\bibfnamefont {D.}~\bibnamefont {Polley}}, \bibinfo {author}
  {\bibfnamefont {N.}~\bibnamefont {Zhou~Hagström}}, \bibinfo {author}
  {\bibfnamefont {S.~S. P.~K.}\ \bibnamefont {Arekapudi}}, \bibinfo {author}
  {\bibfnamefont {A.}~\bibnamefont {Semisalova}}, \bibinfo {author}
  {\bibfnamefont {K.}~\bibnamefont {Lenz}}, \bibinfo {author} {\bibfnamefont
  {B.}~\bibnamefont {Green}}, \bibinfo {author} {\bibfnamefont {J.-C.}\
  \bibnamefont {Deinert}}, \bibinfo {author} {\bibfnamefont {I.}~\bibnamefont
  {Ilyakov}}, \bibinfo {author} {\bibfnamefont {M.}~\bibnamefont {Chen}},
  \bibinfo {author} {\bibfnamefont {M.}~\bibnamefont {Bawatna}}, \bibinfo
  {author} {\bibfnamefont {V.}~\bibnamefont {Scalera}}, \bibinfo {author}
  {\bibfnamefont {M.}~\bibnamefont {d’Aquino}}, \bibinfo {author}
  {\bibfnamefont {C.}~\bibnamefont {Serpico}}, \bibinfo {author} {\bibfnamefont
  {O.}~\bibnamefont {Hellwig}}, \bibinfo {author} {\bibfnamefont {J.-E.}\
  \bibnamefont {Wegrowe}}, \bibinfo {author} {\bibfnamefont {M.}~\bibnamefont
  {Gensch}},\ and\ \bibinfo {author} {\bibfnamefont {S.}~\bibnamefont
  {Bonetti}},\ }\bibfield  {title} {\bibinfo {title} {Inertial spin dynamics in
  ferromagnets},\ }\href {http://dx.doi.org/10.1038/s41567-020-01040-y}
  {\bibfield  {journal} {\bibinfo  {journal} {Nat. Phys.}\ }\textbf {\bibinfo
  {volume} {17}},\ \bibinfo {pages} {245–250} (\bibinfo {year}
  {2020})}\BibitemShut {NoStop}%
\bibitem [{\citenamefont {De}\ \emph {et~al.}(2025)\citenamefont {De},
  \citenamefont {Schlegel}, \citenamefont {Lentfert}, \citenamefont {Scheuer},
  \citenamefont {Stadtmüller}, \citenamefont {Pirro}, \citenamefont {von
  Freymann}, \citenamefont {Nowak},\ and\ \citenamefont
  {Aeschlimann}}]{De_2024}%
  \BibitemOpen
  \bibfield  {author} {\bibinfo {author} {\bibfnamefont {A.}~\bibnamefont
  {De}}, \bibinfo {author} {\bibfnamefont {J.}~\bibnamefont {Schlegel}},
  \bibinfo {author} {\bibfnamefont {A.}~\bibnamefont {Lentfert}}, \bibinfo
  {author} {\bibfnamefont {L.}~\bibnamefont {Scheuer}}, \bibinfo {author}
  {\bibfnamefont {B.}~\bibnamefont {Stadtmüller}}, \bibinfo {author}
  {\bibfnamefont {P.}~\bibnamefont {Pirro}}, \bibinfo {author} {\bibfnamefont
  {G.}~\bibnamefont {von Freymann}}, \bibinfo {author} {\bibfnamefont
  {U.}~\bibnamefont {Nowak}},\ and\ \bibinfo {author} {\bibfnamefont
  {M.}~\bibnamefont {Aeschlimann}},\ }\bibfield  {title} {\bibinfo {title}
  {Magnetic nutation: Transient separation of magnetization from its angular
  momentum},\ }\href {http://dx.doi.org/10.1103/physrevb.111.014432} {\bibfield
   {journal} {\bibinfo  {journal} {Phys. Rev. B}\ }\textbf {\bibinfo {volume}
  {111}},\ \bibinfo {pages} {014432} (\bibinfo {year} {2025})}\BibitemShut
  {NoStop}%
\bibitem [{\citenamefont {Unikandanunni}\ \emph {et~al.}(2022)\citenamefont
  {Unikandanunni}, \citenamefont {Medapalli}, \citenamefont {Asa},
  \citenamefont {Albisetti}, \citenamefont {Petti}, \citenamefont {Bertacco},
  \citenamefont {Fullerton},\ and\ \citenamefont
  {Bonetti}}]{Unikandanunni_2022}%
  \BibitemOpen
  \bibfield  {author} {\bibinfo {author} {\bibfnamefont {V.}~\bibnamefont
  {Unikandanunni}}, \bibinfo {author} {\bibfnamefont {R.}~\bibnamefont
  {Medapalli}}, \bibinfo {author} {\bibfnamefont {M.}~\bibnamefont {Asa}},
  \bibinfo {author} {\bibfnamefont {E.}~\bibnamefont {Albisetti}}, \bibinfo
  {author} {\bibfnamefont {D.}~\bibnamefont {Petti}}, \bibinfo {author}
  {\bibfnamefont {R.}~\bibnamefont {Bertacco}}, \bibinfo {author}
  {\bibfnamefont {E.~E.}\ \bibnamefont {Fullerton}},\ and\ \bibinfo {author}
  {\bibfnamefont {S.}~\bibnamefont {Bonetti}},\ }\bibfield  {title} {\bibinfo
  {title} {Inertial spin dynamics in epitaxial cobalt films},\ }\href
  {http://dx.doi.org/10.1103/physrevlett.129.237201} {\bibfield  {journal}
  {\bibinfo  {journal} {Phys. Rev. Lett.}\ }\textbf {\bibinfo {volume} {129}},\
  \bibinfo {pages} {237201} (\bibinfo {year} {2022})}\BibitemShut {NoStop}%
\bibitem [{\citenamefont {Hartmann}\ \emph {et~al.}(2025)\citenamefont
  {Hartmann}, \citenamefont {Unikandanunni}, \citenamefont {...}, \citenamefont
  {Bonetti},\ and\ \citenamefont {Anders}}]{Hartmann_2025}%
  \BibitemOpen
  \bibfield  {author} {\bibinfo {author} {\bibfnamefont {F.}~\bibnamefont
  {Hartmann}}, \bibinfo {author} {\bibfnamefont {V.}~\bibnamefont
  {Unikandanunni}}, \bibinfo {author} {\bibnamefont {...}}, \bibinfo {author}
  {\bibfnamefont {S.}~\bibnamefont {Bonetti}},\ and\ \bibinfo {author}
  {\bibfnamefont {J.}~\bibnamefont {Anders}},\ }\href@noop {} {\bibinfo {title}
  {\emph{In Preparation.}}} (\bibinfo {year} {2025})\BibitemShut {NoStop}%
\end{thebibliography}%

\newpage
\clearpage
\newpage

\appendix
\section{Ohmic vs. Lorentzian power spectrum}
\label{app:ohmic_vs_Lorentzian_PSD}

\begin{figure*}
    \centering
    \includegraphics[width=0.9\textwidth]{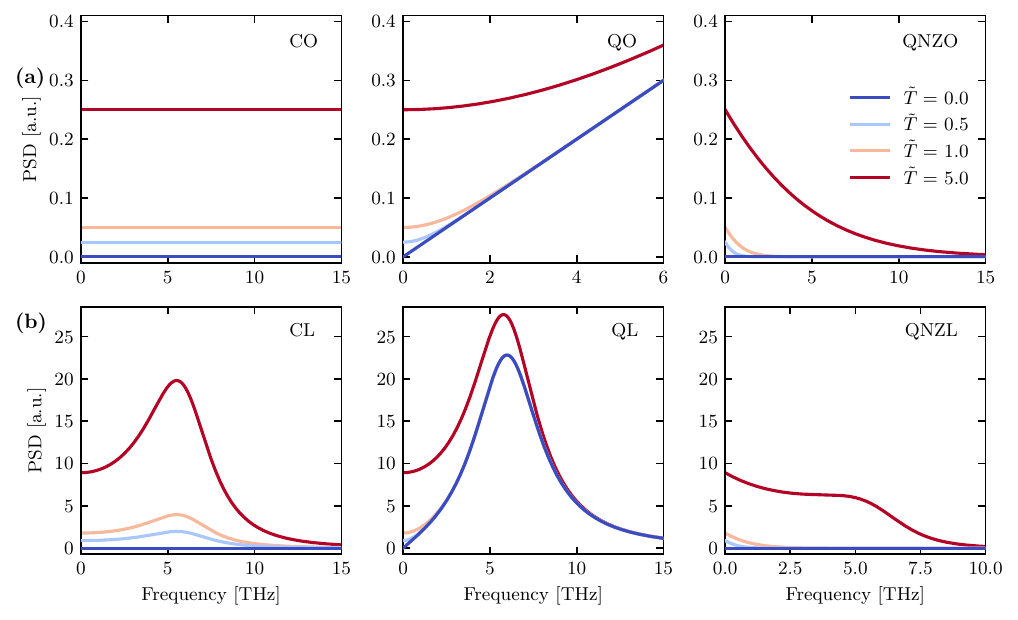}
    \caption{\textbf{Comparison of the Ohmic versus the Lorentzian PSD $\tilde{P}(\omega)$:} We here compare the six different power spectra (cf. Tab.~\ref{tab:taxonomy}). (a) Ohmic spectral density (CO, QO, QNZO) and (b) Lorentzian spectral density (CL, QL, QNZL from left to right).
    This demonstrates that the choice of spectral density and of the thermostat is crucial. 
    The effective (unit-free) temperature $\tilde{T}$ is given by  $k_{\mathrm{B}}T/\hbar\tilde{\omega}_{\mathrm{L}}$. The Ohmic spectral density is given by $I^{\mathrm{Ohm}}(\omega) \propto \alphaG\omega$.
    The Lorentzian parameters for the Lorentzian PSD's are those for Ni (see Tab.~\ref{tab:parameters}).
    }
    \label{fig:ohmic_vs_Lorentzian}
\end{figure*}

In Fig.~\ref{fig:ohmic_vs_Lorentzian} we compare the power spectra of an (a) Ohmic and (b) Lorentzian spectral density (Ni parameters, see Tab.~\ref{tab:parameters}) with each of the three thermostats, i.e. classical, quantum, quantum-no-zero~\eqref{eq:powerspectrum} (see also Tab.~\ref{tab:taxonomy}).
The Ohmic spectral density is given by $I^{\mathrm{Ohm}}(\omega) \propto \alphaG\omega$ and is commonly used in ASD simulations, e.g.~\cite{Evans_2014,Barker_2019}.
Each PSD is plotted for four effective temperatures $\tilde{T}$ and shows how relevant the choice of the spectral density and the thermostat is.

It is helpful to discuss the relevant qualitative features of the choice of thermostat and spectral density on the power spectrum. For classical Ohmic noise (CO) in Fig.~\ref{fig:ohmic_vs_Lorentzian}(a) the PSD is flat and simply increases in magnitude with increasing temperature. This means that there is no frequency bias to the excited fluctuations and therefore the noise has no spectral characteristics. This allows for high frequency excitations to occur readily, but also gives a classical scaling of the noise with temperature leading to a classical temperature-dependent magnetization curve that is not accurate for most magnetic materials. The classical thermostat with Lorentzian (CL) noise in Fig.~\ref{fig:ohmic_vs_Lorentzian}(b) introduces a spectral character to the effective PSD, that promotes resonant thermal excitations at $\omega = \omega_0$, while depopulating the available excitations at higher frequencies $\omega > \omega_0$. Since $\omega_0$ is at a relatively high frequency and significantly above the precession frequency $\tau_\mathrm{p} =  ~\gamma \alphaG \sim 6 \times 10^{10}$ Hz, resonant effects only typically occur under ultrafast (sub-picosecond) excitation. Correspondingly, the spin excitations relevant for the equilibrium magnetization are all below $\omega_0$ and so the dominant effect is from the thermostat where again the PSD scales linearly with the temperature. 

Introducing the quantum thermostat with Ohmic noise (QO) in Fig.~\ref{fig:ohmic_vs_Lorentzian}(c) leads to a non-linear scaling of the PSD amplitude with temperature, as well as a linear increase in the PSD as a function of frequency. 
This naturally leads to fluctuations at zero temperature due to remaining fluctuations of quantized phonon modes. The non-linear scaling of the PSD with temperature is reminiscent of the qualitative effect of spin-temperature rescaling~\cite{Evans_2015,Berritta_2024} and leads to a non-classical temperature-dependent magnetization, but within a quantum mechanical framework. The use of the Lorentzian PSD with quantum thermostat (QL) noise in Fig~\ref{fig:ohmic_vs_Lorentzian}(d) combines the spectral features described for the classical thermostat above, with the non-linear scaling of the noise with temperature and zero-point motion. 

It is possible to remove the zero-point motion effects of the quantum noise by using the thermostat in Eq.~\eqref{eq:barker}. Here the quantum no-zero thermalized PSD with Ohmic noise (QNZO) in Fig.~\ref{fig:ohmic_vs_Lorentzian}(e) decays rapidly in frequency, leading to a strong depopulation of high frequency modes. This limits spin wave excitations to low-frequencies and can therefore significantly increase the Curie temperature depending on the choice of $S_0$. Finally in the case of quantum no-zero PSD with Lorentzian noise in Fig.~\ref{fig:ohmic_vs_Lorentzian}(f) some spectral character is introduced into the noise, but only when the temperature is large enough that characteristic thermal excitation overlaps with $\omega_0$. At low temperatures the QNZO and QNZL should exhibit very similar behavior, while at high temperatures the Lorentzian nature of the noise allows for the high frequency spin wave modes characteristic of the phase transition, and therefore does not increase the Curie temperature as much as the QNZO case. As before for the CL noise, with sufficient temperature the QNZL should exhibit spectral characteristics with non-Markovian behavior.

\section{Fitting the (phonon) DOS data}
\label{app:fitDOS}
In our implementation of the \mLLG~equation the Lorentzian parameters $(\alp,\omega_0,\Gamma)$ in units of $\mathrm{rad}\cdot\mathrm{THz}$ are transformed to unitless variables via  $\tilde{\alp} = \alp/\tilde{\omega}_{\mathrm{L}}^3$, $\tilde{\omega}_0 = \omega_0/\tilde{\omega}_{\mathrm{L}}$, and $\tilde{\Gamma} = \Gamma/\tilde{\omega}_{\mathrm{L}}$, are scaled via an \textit{effective} Larmor frequency  $\tilde{\omega}_{\mathrm{L}}$ which is defined for a fixed field strength $|\mathbf{B}| = 1.0$~T. Thus, $\tilde{\omega}_{\mathrm{L}} = |\gamma|\cdot1\,\mathrm{T} \approx 0.176\,\mathrm{rad}\cdot\mathrm{THz}$, with $|\gamma| \approx 1.76\cdot10^{11}\,$rad/s/T being the electron gyromagnetic ratio.
The \textit{effective} Larmor frequency $\tilde{\omega}_{\mathrm{L}}$ is solely used to transfer the dimensionless simulation parameters into dimensioned parameters and vice versa~\cite{Scali_2024}.

\begin{figure*}
    \centering
    \includegraphics[width=0.9\textwidth]{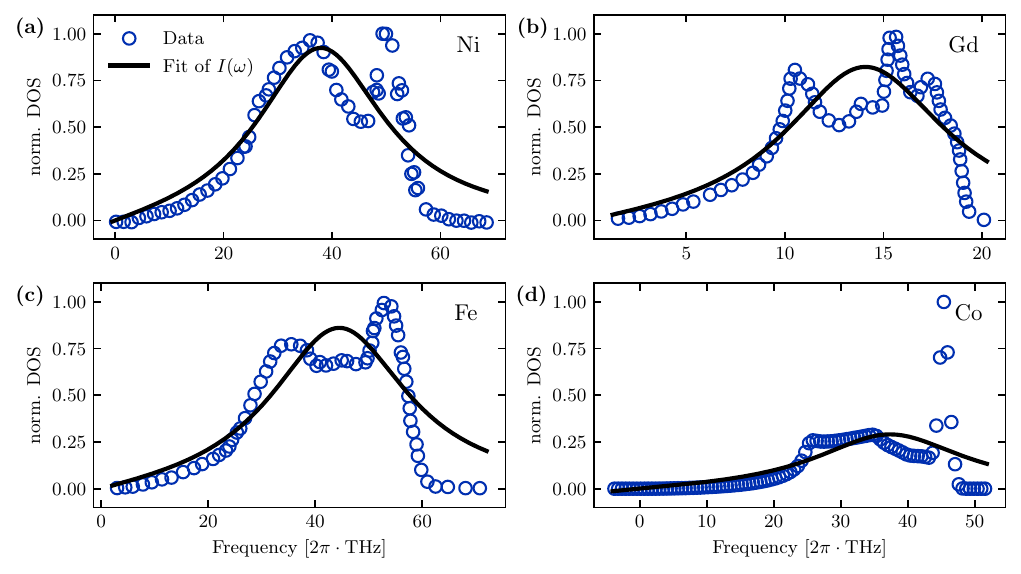}
    \caption{\textbf{Lorentzian fitting of the phonon DOS for four materials:} The fitted parameters are (a) Ni $\omega_0 = 40.5\,\mathrm{rad}\cdot\mathrm{THz}$ and $\Gamma = 28.9\,\mathrm{rad}\cdot\mathrm{THz}$, (b) Gd $\omega_0 = 14.9\,\mathrm{rad}\cdot\mathrm{THz}$ and $\Gamma = 10.0\,\mathrm{rad}\cdot\mathrm{THz}$, (c) Fe $\omega_0 = 47.1\,\mathrm{rad}\cdot\mathrm{THz}$ and $\Gamma = 31.5\,\mathrm{rad}\cdot\mathrm{THz}$, and (d) Co $\omega_0 = 39.7\,\mathrm{rad}\cdot\mathrm{THz}$ and $\Gamma = 27.1\,\mathrm{rad}\cdot\mathrm{THz}$. 
    The parameters in units of THz are given in Tab.~\ref{tab:parameters}.
    }
    \label{fig:DOS_fits}
\end{figure*}

\section{Comparing the paramagnetic dynamics}
\label{app:para_dynamics}

To validate our implementation of the open-system Landau-Lifshitz-Gilbert (\mLLG) equation with colored noise in Vampire, we first examined the paramagnetic case ($J=0$) through comparison with \textsc{SpiDy}~\cite{Scali_2024}.

\begin{figure*}
    \centering
    \includegraphics[width=0.9\textwidth]{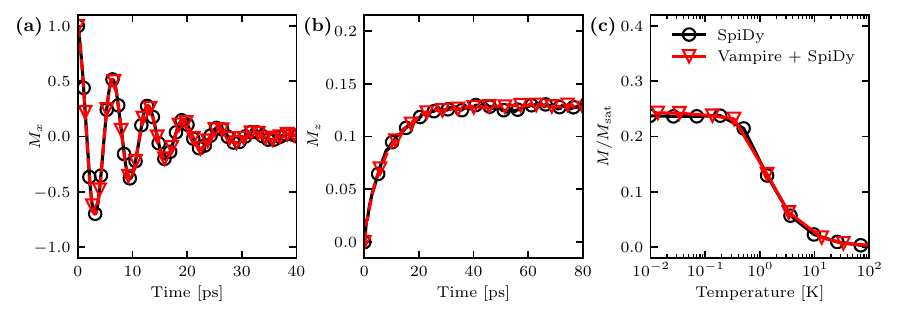}
    \caption{\textbf{Paramagnetic magnetization dynamics:} Here we compare the paramagnetic spin dynamics calculated from SpiDy~\cite{Scali_2024} (black line, open circles) and the implementation of the \mLLG~equation~\eqref{eq:mLLG} into Vampire (red line, open triangles), while employing a quantum thermostat $\tilde{\lambda}_{\mathrm{qu}}$~\eqref{eq:quantum}. 
    (a)-(b) We find excellent agreement between the two approaches, here exemplary shown for the $M_x$ and $M_z$ components of the magnetization. 
    (c) This leads to an excellent agreement between the steady-states of the two implementations over the complete temperature range.
    The Lorentzian simulation parameters are taken from Set 1 of Ref.~\cite{Anders_2022}, i.e. $\tilde{\alp} = 10.0~\tilde{\omega}_{\mathrm{L}}^3$, $\tilde{\omega}_0 = 7.0~\tilde{\omega}_{\mathrm{L}}$, and $\tilde{\Gamma} = 5.0~\tilde{\omega}_{\mathrm{L}}$, with $\tilde{\omega}_{\mathrm{L}} = 0.176\,\mathrm{rad}\cdot\mathrm{THz}$.
    Further, the spin length of the classical spin vector is $S_0 = \hbar/2$, the external field is $\mathrm{H}_{\mathrm{ext}} = (0,0,1)$T. For (a)-(b) the single spin temperature is $T = 1.35$~K and the dynamics is averaged over $50000$, respectively, over $13500$ spin trajectories for panel (c).
    The spin is initial aligned along the positive $x$-direction, i.e. $\mathbf{S}_0 = (1.0,0.0,0.0)$.
    }
    \label{fig:para_dyn}
\end{figure*}

Fig.~\ref{fig:para_dyn}(a)-(b) demonstrates excellent agreement between the dynamics of \textsc{SpiDy} (black line, open circles) and our \mLLG \textsc{Vampire} implementation (red line, open triangles).  
In addition, Fig.~\ref{fig:para_dyn}(c) shows perfect agreement between the paramagnetic steady-state of the two approaches.
The equilibrium magnetization was determined by evolving the system under a fixed quantum thermostat (see Eq.~\eqref{eq:quantum}) until the magnetization amplitude $\mathrm{m}(t) = |\langle \mathbf{m}(t)\rangle|$ reached steady state. For each temperature the final magnetization amplitude was computed by averaging over a time window in the steady-state regime.

To compare the accuracy in more detail, we ran paramagnetic simulations with a classical thermostat~\eqref{eq:classical} with $T=0$~K. 
The system of coupled differential equations~\eqref{eq:generalizedLLG_numerI}-\eqref{eq:generalizedLLG_numerIII} becomes deterministic and the accuracy is not falsified by a limited number of stochastic realizations.
We find that the absolute error (AE) between the two approaches is $|\mathrm{AE}| < 6\cdot10^{-6}$.
Therefore, we can use our implementation of \textsc{SpiDy} into \textsc{Vampire} to study magnetic materials beyond paramagnets.

\section{Curie temperature for the different thermostats}
\label{app:Curie_temperature}
\begin{figure*}
    \centering
    \includegraphics[width=0.48\textwidth]{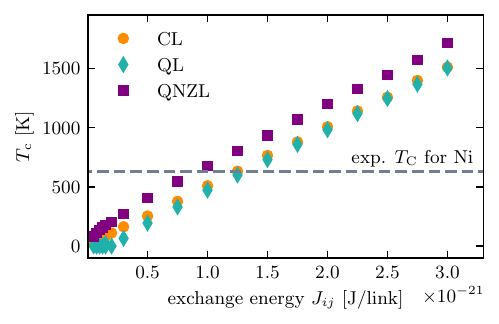}
    \caption{\textbf{Curie temperature of the three different thermostats for varying exchange energies:} We vary the exchange energy $J_{ij}$ and determine the corresponding Curie temperature $T_{\mathrm{C}}$. 
    The classical (orange circle), quantum (green diamonds), and quantum-no-zero (purple squares) thermostats show the same linear trend for $J_{ij} > 0.3\times10^{-21}$~J/link.
    However, close attention should be paid to the slopes of the three thermostats. The classical and quantum thermostat have a similar slope, whereas the slope of the quantum-no-zero noise is steeper.
    This implies that the prediction of the Curie temperature and the exchange energy is strongly related to choice of the thermostat.
    We choose the same Lorentzian parameters as for Ni, compare to Tab.~\ref{tab:parameters}. 
    The experimentally measured Curie temperature for Ni ($T_{\mathrm{C}} = 631.2$~K) is indicated by the gray dashed line. In classical ASD simulations with \textit{Vampire} an exchange energy of $J_{ij} = 6.0\times10^{-21}$ J/link reproduces the correct $T_{\mathrm{C}}$~\cite{Evans_2014}.
    }
    \label{fig:TcvsJ}
\end{figure*}

In Fig.~\ref{fig:TcvsJ} we demonstrate that the choice of thermostat has a strong influence on the numerically observed Curie temperature $T_{\mathrm{C}}$ of the \mLLG equation~\eqref{eq:mLLG} framework. 
We find, for the Lorentzian parameters of Ni and a given exchange energy $J_{ij}$, that the QNZL thermostat consistently predicts a larger $T_{\mathrm{C}}$ than the CL and QL thermostats. 
This is directly linked to the discussion of the different PSD's in Appendix~\ref{app:ohmic_vs_Lorentzian_PSD}.
In contrast, in classical ASD simulations with \textsc{Vampire}, using white noise, an exchange energy of $J_{ij} = 6.0\times10^{-21}$ J/link has been reported to reproduces the correct $T_{\mathrm{C}}$~\cite{Evans_2014}. 
It follows that the thermostat quite drastically changes the observed Curie temperature.
Future work is needed to gain a detailed understanding of all its implications. 

\section{Non-unity magnetization for QL noise}
\label{app:Quantum_noise_disc}
QL noise naturally leads to zero-point fluctuations which significantly reduces the magnetization at low temperatures rendering our QL noise simulations unpyhsical, compare with Fig.~\ref{fig:equilibrium_noise}(e). This however could not be a problem of the theroy, but rather the model to simulate the equilibrium magnetization as the spins in ASD are treated classically which overestimated the effect of zero-point fluctuations. 

Additionally zero-point fluctuations of atomic spin moments raises an interesting and important question with regard to the electronic structure of a magnetic material in the presence of a phononic heat bath. In standard electronic structure calculations such as density functional theory (DFT) a time-independent solution is assumed, where local fluctuations of nuclear (atomic) positions are averaged out, the electron density is then minimized and the local spin moment is calculated, typically with good estimations of the local moment value in comparison with experimental data. In some respects, the quantum spin fluctuations are included within the electronic structure, projecting the net magnetic moment onto a classical vector which is the relevant observable quantity. On the other hand, small zero-point fluctuations of the lattice will naturally induce spin fluctuations due to small changes in the electronic structure. Within the framework of the open-system LLG, the strength of these quantum fluctuations is determined by the parameter $S_0$, which for correspondence with the classical case is assumed the same as the local atomic spin moment, $S_0 := \mu_i$. However, this assumes that the strength of the fluctuations in the case of QL noise is based on the small and finite length of the spin moment $\mu_i$, rather than some auxiliary quantity such as the mass of the atom (that determines the strength of the zero-point motion). In addition quantum fluctuations of the electronic bath itself are not considered here, nor in DFT calculations of electronic structure and may also add additional sources of noise. It is possible to separate the values of $S_0$ (which governs the strength of the fluctuations) and $\mu_i$ (which governs the strength of coupling to an external magnetic field) through an $s-d$ model that explicitly considers local and itinerant magnetic moments, but that is somewhat beyond the scope of the present article as it requires a more detailed understanding of the physical nature of the thermal bath.  

\section{Symbols}
See Tab.~\ref{tab:symbols}.

\begin{table*}[h!]
\centering
\caption{\textbf{List of symbols} used in the paper and their meanings (excluding standard and redundant definitions).}
\begin{tabular}{lc}
\hline
\hline
\textbf{Symbol} & \textbf{Description} \\
\hline
$K(t - t')$ & Memory kernel capturing non-Markovian effects \\
$\tilde{P}(\omega, T)$ & Power spectral density (PSD) of the thermal noise \\
$I(\omega)$ & Spectral density of the bath (Ohmic or Lorentzian) \\
$\Theta(\omega, T)$ & Thermostat determining temperature dependence of noise \\
$\omega_0$ & Resonance (center) frequency of the Lorentzian bath \\
$\Gamma$ & Width of the Lorentzian spectral peak (broadening) \\
$\alp$ & Coupling strength between spin system and bath (spin-orbit coupling) \\
$\tilde{\omega}_\mathrm{L}$ & Effective Larmor frequency: $\tilde{\omega}_L = |\gamma| \cdot 1\,\text{T}$ \\
$\mathbf{V}_i, \mathbf{W}_i$ & Auxiliary variables for memory effects in numerical integration \\
\hline
\hline
\end{tabular}
\label{tab:symbols}
\end{table*}

\cleardoublepage 

\end{document}